\documentclass[11pt,a4paper]{article}
\usepackage {amsmath, amssymb, amsfonts}
\usepackage {graphics}
\usepackage {graphicx}
\usepackage {longtable}
\usepackage {pgf,tikz,url}
\usepackage {float,afterpage,color} 
\usepackage {verbatim} 
\usetikzlibrary{arrows,automata,shapes,backgrounds,topaths,petri}
\usepackage {subfigure}
\usepackage{chemarr} 
\usepackage{lscape}
\usepackage{setspace}
\usepackage{fancyheadings}
\usepackage{float,afterpage}
\usepackage{bibspacing}
\usepackage{multirow}

\usepackage{anysize}
\marginsize{2cm}{2cm}{2cm}{2cm}

\renewcommand\Re{\operatorname{Re}}
\renewcommand\Im{\operatorname{Im}}
\DeclareMathOperator{\argmax}{argmax}
\DeclareMathOperator{\argmin}{argmin}

\usepackage{booktabs} 
\usepackage{array} 
\usepackage{paralist} 
\usepackage{verbatim} 
\usepackage{subfigure} 

\usepackage{fancyhdr} 
\usepackage{sectsty}
\allsectionsfont{\sffamily\mdseries\upshape} 

\begin{document}

\title{\bf Bayesian design of synthetic biological systems}
\author{\sc Chris Barnes, Daniel Silk, Xia Sheng and Michael P.H. Stumpf
\\[3mm] Theoretical Systems Biology Group, Division of Molecular Biosciences,\\ Imperial College London, London SW7 2AZ, UK\\[2mm] \url{http://www.theosysbio.bio.ic.ac.uk}\\[3mm] \small Email: christopher.barnes@imperial.ac.uk, m.stumpf@imperial.ac.uk}
\date{}
\maketitle

\begin{abstract}
Here we introduce a new design framework for synthetic biology that exploits the advantages of Bayesian model selection. We will argue that the difference between inference and design is that in the former we try to reconstruct the system that has given rise to the data that we observe, while in the latter, we seek to construct the system that produces the data that we would like to observe, i.e. the desired behavior. Our approach allows us to exploit methods from Bayesian statistics, including efficient exploration of models spaces and high-dimensional parameter spaces, and the ability to rank models with respect to their ability to generate certain types of data. Bayesian model selection furthermore automatically strikes a balance between complexity and (predictive or explanatory) performance of mathematical models. In order to deal with the complexities of molecular systems we employ an approximate Bayesian computation scheme which only requires us to simulate from different competing models in order to arrive at rational criteria for choosing between them. We illustrate the advantages resulting from combining the design and modeling (or {\em in-silico} prototyping) stages currently seen as separate in synthetic biology by reference to deterministic and stochastic model systems exhibiting adaptive and switch-like behavior, as well as bacterial two-component signaling systems. 
\end{abstract}

\section{Introduction}
As we are beginning to understand the mechanisms governing biological systems we are starting to identify potential ways of guiding or controlling the behavior of cellular and molecular systems. Rationally reengineering organisms for biomedical or biotechnological purposes has become the central aim of the fledgling discipline of synthetic biology. By redirecting regulatory and physical interactions or by altering molecular binding affinities we may, for example, control metabolic processes \cite{Martin:2003p2607, Ro:2006p2764} or alter intra and inter cellular communication and decision making processes \cite{You:2004p4217,Kobayashi:2004p6133}. The range of potential applications of such engineered systems is vast: designing microbes for biofuel 
production \cite{Fortman:2008p2396,Savage:2008p2620} and bioremediation \cite{Cases:2005p2750}; developing control strategies which drive stem cells through the various decisions to become terminally differentiated 
(or back) \cite{Takahashi:2007p6160,Hanna:2010p6205}, with the aim of developing novel therapeutics \cite{Lu:2009p6163, Macarthur:2009p6166}; 
construction of new drug-delivery systems with homing microbes delivering molecular medicines 
directly to the site where they are needed \cite{Anderson:2006p2634}; use of bacteria or bacterial populations (employing 
swarming and quorum sensing) as biosensors \cite{Rajendran:2008p6134}; and gaining better understanding of all manner 
of biological systems by systematically probing their underlying molecular machinery. 
\par
A range of tools and building blocks for such engineered biological systems are now available 
which allow us to, at least in principle, build such systems from simple and reusable biological components \cite{Canton:2008p2395}. In electronic systems, such modularity has been crucial and has allowed the cost-effective production of reliable components that can be combined to produce desired outputs. 
Biology, however, poses different and novel challenges that are intimately linked to the biophysical 
and biochemical properties of biomolecules and the media in which they are suspended. Especially in crowded environments such as found inside living cells the lack of insulation between different components, i.e. the very real possibility of undesired cross talk, can create problems; 
with increasing miniaturization similar, albeit quantum effects, are now also surfacing in electronic circuits \cite{Liang:2002p6209}. 
\begin{figure}[htb]
\begin{center}
\centerline{ \includegraphics[width=0.5\textwidth]{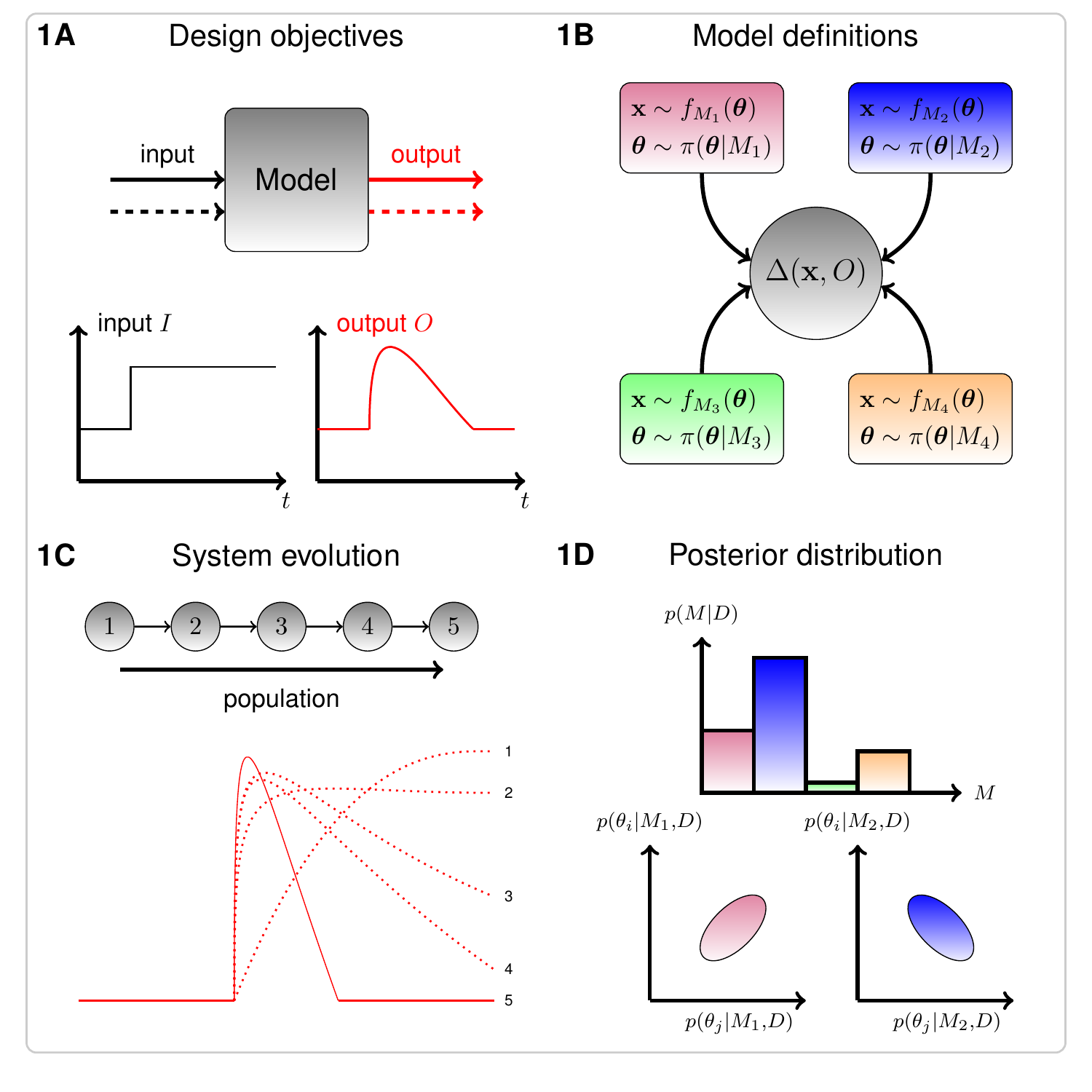} }
\caption{Bayesian approach to system design. A) The design objectives are encoded by the specification of input and output characteristics. B) One or more competing designs for the system are specified together with priors on the parameters. A distance function, $\Delta( \mathbf{x},O)$, relates model output to the desired output characteristic. C) The system is evolved using sequential Monte Carlo. Each population more accurately approximates the desired behavior. D) The model posterior probability encodes the ability of each design to achieve the desired behavior. The parameter posterior shows parameters that are sensitive or insensitive to the input-output specification.   }
\label{fig:method1}
\end{center}
\end{figure}
\par
As synthetic biology gears up to bring engineering methods and tools to bear on biological 
problems the way in which we manipulate biological systems and processes is likely to change. Historically, each new branch of engineering has gone through a phase of what can be described as tinkering before rationally planned and executed designs became common place. Arguably, this is the current state of synthetic biology and it has indeed been suggested that the complexity of synthetic biological systems over the past decade has reached a plateau \cite{Purnick:2009p2394}. From the earliest days, explicit quantitative 
modeling of systems has been integral to the vision and practice of synthetic biology and it will become increasingly important in the future. The ability 
to model how a natural or synthetic system will perform under controlled conditions must in fact be seen as one of the hallmarks of success of the integrative approaches taken by systems and synthetic biology. 
\par
Here we present a statistical approach to the design of synthetic biological systems that utilizes methods from Bayesian statistics to train models according to specified input-output characteristics. It incorporates modeling and automated design and is general in the sense that it can be applied to any system that can be described by a mathematical model which can be simulated. Because of the statistical nature of this approach, previously challenging problems such as handling stochastic models, accounting for kinetic parameter uncertainty and incorporating environmental stochasticity can all be handled in a straightforward and consistent manner.

\section{Bayesian approach to system design}
The question of how to design a system to perform a specified task can be viewed as an analogue to reverse engineering. In design we want to elucidate the most appropriate system to achieve our design objectives; in reverse engineering we aim to infer the most probable system structure and dynamics that can give rise to some observational data. In this respect, the design question
can be viewed as statistical inference on data {\em we wish to observe}.
\par
In the Bayesian approach to statistical inference the posterior distribution is the quantity of interest and this is given by the normalized product of the {\em likelihood} and the {\em prior}. In most practical applications the posterior distribution cannot be derived analytically but if the likelihood (and prior) can be expressed mathematically we can use Monte Carlo methods to sample from the posterior. In many cases where the model structure is complex the likelihood cannot be written in closed form and traditional Monte Carlo techniques cannot be applied. These include inference for the types of stochastic processes encountered in systems and synthetic biology. In these cases a family of techniques known collectively as approximate Bayesian computation (ABC) can be applied: these use model simulations to approximate the posterior distribution directly. Here we use a sequential Monte Carlo ABC algorithm known as ABC SMC to move from the prior to the approximate posterior via a series of intermediate distributions \cite{Toni:2009p606}. This framework can also be used to perform Bayesian model selection \cite{Toni:2010p2027} and has been implemented in the software package ABC-SysBio \cite{Liepe:2010p2858}. 
\par
Figure \ref{fig:method1} depicts the approach presented here. The design objectives are first specified through input-output characteristics. Here these have been depicted as a single time series, though the method can be applied in a much broader sense with multiple inputs and outputs. A set of competing designs is then specified through deterministic or stochastic models each containing a set of kinetic parameters and associated prior distributions. The distance function measures the discrepancy between the model output and the objective. In principal it is possible to specify a distribution over the objective and each model could also contain experimental error.  The ABC SMC algorithm then automatically evolves the set of models towards the desired design objectives. The results are a set of posterior probabilities representing the probability for each design to achieve the specified design objectives in addition to the posterior probability distribution of the associated kinetic parameters. This approach is similar in spirit to some existing methods for the automated design of genetic networks such as those adopting evolutionary algorithms \cite{Bray:1994p3254,Francois:2004p2367}, Monte Carlo methods \cite{Battogtokh:2002p3179, Feng:2004p3168} or optimization \cite{Rodrigo:2007p2375, Dasika:2008p2374,Batt:2007p2392} but the advantages of our method over traditional ones are that we can utilize powerful concepts from Bayesian statistics in the design of complex biological systems, including
\begin{itemize}
\item the rational comparison of models under parameter uncertainty using Bayesian model selection which automatically accounts for model complexity (number of parameters) and robustness to parameter uncertainty
\item a posterior {\em distribution} over possible design parameter values that can be analyzed for parameter sensitivity and robustness and provide credible limits on design parameters
\item the treatment of stochastic systems at the {\em design} stage including the design of systems with required {\em probability distributions} on system components.
\item methods for the efficient exploration of high dimensional parameter space
\end{itemize}
\par
In the following we demonstrate the power of this approach by examining, from this new perspective, systems that have been of interest in the recent literature. First we consider systems that are capable of biochemical adaptation \cite{Ma:2009p2043}. We then look at the ability of two bacterial two component system (TCS) topologies to achieve particular input-output behaviors; and finally we finish with an analysis of designs for a stochastic toggle switch with no cooperative binding at the promoter. 

\section{Biochemical adaptation}
\begin{figure*}[t]
\begin{center}
\centerline{ \includegraphics[width=1.0\textwidth]{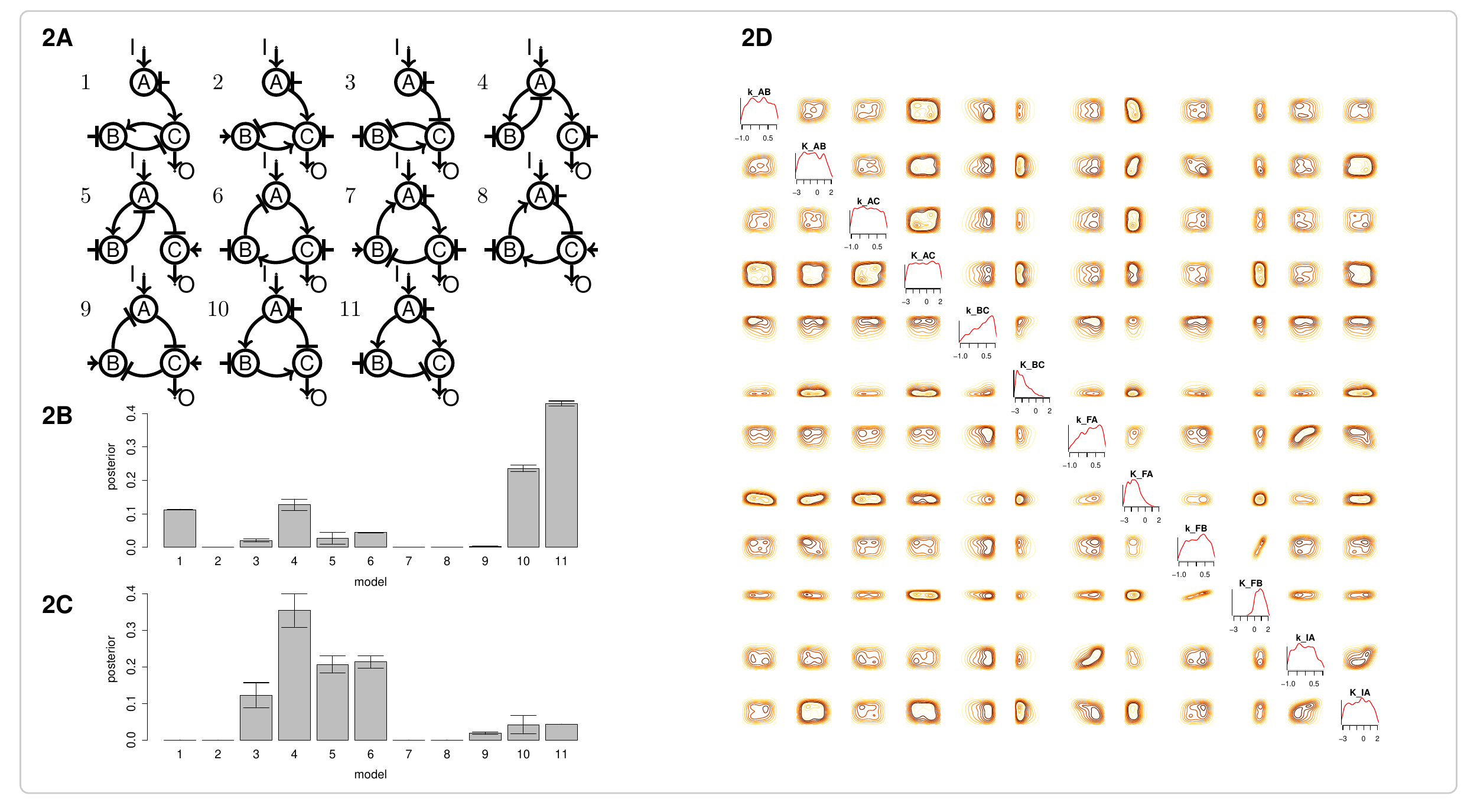} }
\caption{Biochemical adaptation. A) 11 networks capable of biochemical adaptation. $A,B,C$ represent enzymes that catalyze reactions in their active state. For example $A \rightarrow B$ indicates that $A$ converts $B$ from its inactive to active state and $A \dashv B$ indicates that $A$ converts $B$ from its active to inactive state. The input is applied to species $A$ and the output is taken to be the concentration of the active form of $C$. The concentrations of active and inactive forms sum to 1. Reactions with no origin node refer to background activating/deactivations enzymes. B) Posterior probability for achieving biochemical adaptation when there is no cooperativity. C) Posterior probability for achieving biochemical adaptation when cooperativity is included. The error bars in panels B and C indicate the variability in the marginal model posteriors over three separate runs. D) Parameter posterior distribution, represented by univariate and bivariate marginal distributions, for model 11 in the case of no cooperativity. }
\label{fig:adapt1}
\end{center}
\end{figure*}
Biochemical adaptation refers to the ability of a system to respond to an input signal and return to the pre-stimulus steady state (Figure \ref{fig:method1}A). Ma {\em et al} \cite{Ma:2009p2043} identified two three-node network topologies that are necessary for biochemical adaptation: a negative feedback loop with a buffering node (NFBLB) and an incoherent feedforward loop with a proportioner node (IFFLP). Within these categories they identified eleven simple networks that were capable of adaptation (Figure \ref{fig:adapt1}A). We applied the Bayesian design approach to these eleven networks using Michaelis-Menten kinetic models with and without cooperativity (see appendix B for the ordinary differential equations (ODEs) describing these models). The desired output characteristics were defined through the adaptation efficiency, $E$, and sensitivity, $S$, given by
\begin{align*}
E = \Bigg\lvert \frac{ (O_2 - O_1)/O_1 }{ (I_2 - I_1)/I_1}  \Bigg\rvert \nonumber & \; S = \Bigg\lvert \frac{ (O_{peak} - O_1)/O_1 }{ (I_2 - I_1)/I_1} \Bigg\lvert,
\end{align*}
where $I_1, I_2$ are the input values (here fixed at 0.5 and 0.6 respectively), $O_{1}, O_{2}$ are the output steady state levels before and after the input change and $O_{peak}$ is the maximal transient output level. We defined the two component distance to be $\epsilon = \{E,S^{-1} \}$ such that as $\epsilon$ decreases the behavior approaches the desired behavior. The final population was defined to obey the toletances $\epsilon = \{0.1,1.0\}$, which defines close to perfect adaptation (when $O_{1} -O_{2} \le O_{1}/50$) and a fractional response equal to the fractional change in input.
\par
The results of the model selection are shown in Figure \ref{fig:adapt1} (B and C). When cooperativity is not included the most robust designs for producing the desired input-output characteristics are the incoherent feedforward loops, but when cooperativity is added the posterior shifts significantly towards the negative feedback topology. If a system with these requirements were to be implemented then not only would designs 11 and 4 be clear candidates for further study, but many of the designs can be effectively ruled out and the ranking of the models provides a clear strategy for an experimental programme. These results also illustrate how small changes in context or incomplete understanding of a system can produce a large change in the most robust design. The Bayesian framework allows us to incorporate such uncertainty --- or safeguard against our ignorance --- naturally into the design process.
\par
The posterior distribution provides information on which parameters are correlated and which are the most sensitive to the desired behavior. The posterior for model 11 under no cooperativity is shown in Figure \ref{fig:adapt1}D, where the ODE model is given by
\begin{align*}
\frac{dA}{dt} &=  I k_{IA} \frac{(1-A) }{(1-A) + K_{IA} } - F_{A} k_{FA}\frac{ A }{A + K_{FA} }\\
\frac{dB}{dt} & = A k_{AB} \frac{(1-B) }{(1-B) + K_{AB} } - F_B k_{FB}\frac{B }{B + K_{FB} }\\
\frac{dC}{dt} & = A k_{AC}\frac{(1-C) }{(1-C) + K_{AC} } - B k_{BC} \frac{C }{C + K_{BC} },
\end{align*}
where $X=\{A,B,C\}$ corresponds to the concentrations of the active forms of proteins ($1-X$ corresponds to the concentrations of the inactive form). $I$ represents the input signal and the $k$ and $K$ represent the reaction rate parameters (of which there are 12 in total). $F_A$ and $F_B$ represent background deactivating enzymes with concentration fixed to 0.5.
\par
The posterior shows in particular that the parameters for the background deactivating enzyme ($k_{FB}$ and $K_{FB}$) on node B are correlated, and that $K_{FB}$ should be large; this is exactly the requirement for the linear regime necessary for the IFFLP system to achieve adaptation \cite{Ma:2009p2043}. A principal component analysis of the posterior (Figure S1) shows other correlated parameters on the first few principal components. The last principal component describes the direction of least variance and therefore the most sensitive parameters. From this we can deduce for example that the reaction between nodes A and C is relatively unimportant. A similar analysis for model 4 in the case of cooperativity (Figures S2 and S3) shows for example that the behavior is insensitive to the values of the Hill coefficients and the details of the reaction between nodes A and C.

\section{Robust oscillator design}
\begin{figure}[ht]
\begin{center}
\centerline{ \includegraphics[width=0.5\textwidth]{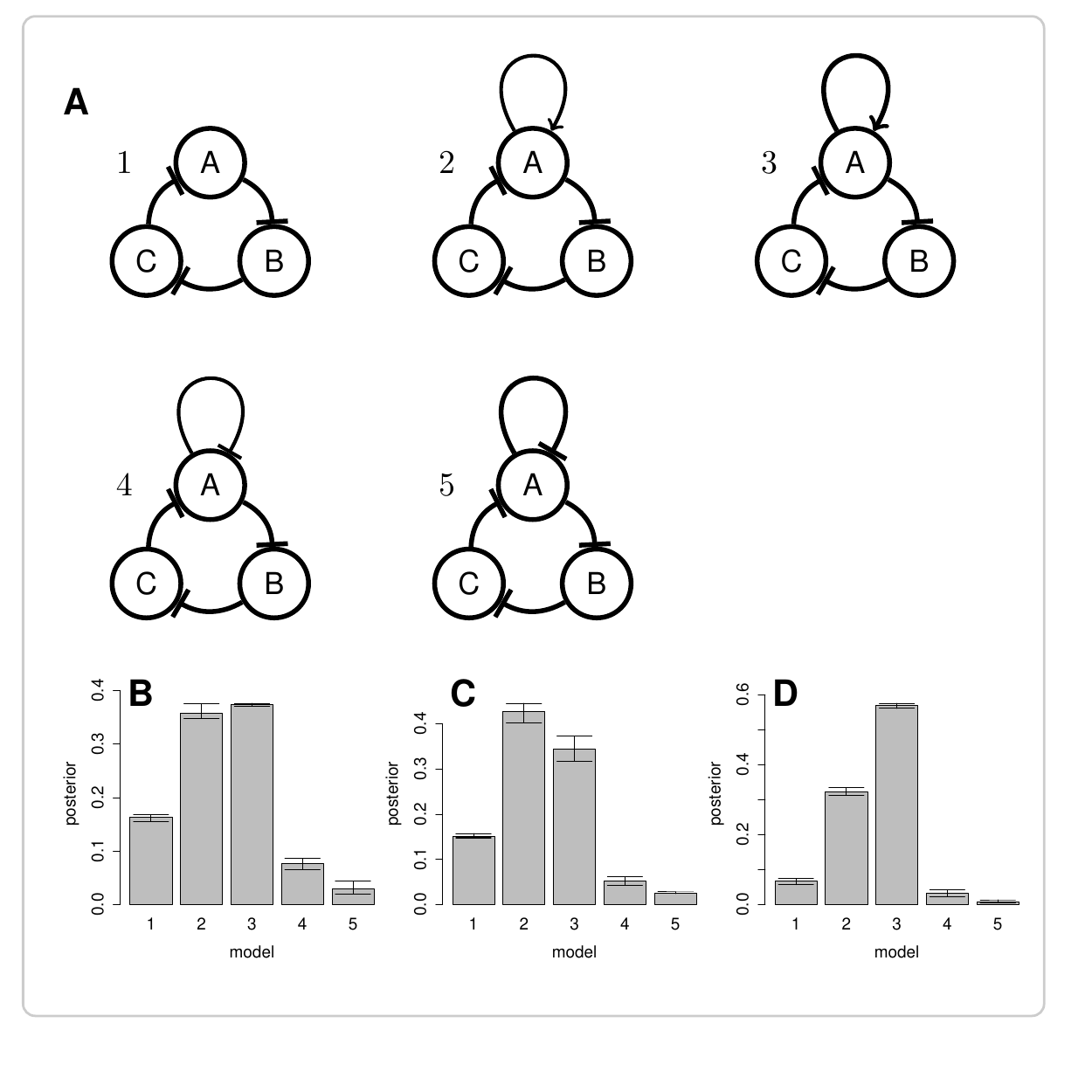}}
\caption{Robust oscillator models. A) 5 oscillator models. Model 1 is a loop of repressive enzymatic reactions. Models 2 and 3 have an additional positive feedback loop on node A with the feedback strength stronger in model 3 (represented by the thicker loop). Models 4 and 5 have an additional negative feedback loop on node A with the feedback strength stronger in model 5. B) Posterior probability for achieving Hopf bifurcation type limit cycle oscillations. C) Posterior probabilities for species A achieving oscillations with amplitude of 0.1 and a frequency of 1Hz. D) Posterior probabilities for species C achieving oscillations with amplitude of 0.1 and a frequency of 1Hz. The error bars in panels B, C and D indicate the variability in the marginal model posteriors over three separate runs.}
\label{fig:osc1}
\end{center}
\end{figure}

Biochemical oscillations are increasingly being implemented in various synthetic systems \cite{Elowitz:2000p2700,Stricker:2008p1958,Tigges:2009p2143,Purcell:2010p4927}. A recent study by Tsai {\em et al} \cite{Tsai:2008p56} compared the ability of five small networks to achieve oscillations. The five designs are shown in Figure \ref{fig:osc1}A where each node represents the active form of a protein, edges represent enzymatic reactions and thicker edges represent increased feedback strength. We applied our Bayesian design methodology to the original problem and further investigated the ability of these designs, provided in detail in appendix C, to achieve particular amplitude-frequency values. 
\par
Figure \ref{fig:osc1}B shows the posterior probability for each model to achieve limit cycle behavior induced by a Hopf bifurcation. The addition of the negative feedback loop in models 4 and 5 does not improve the ability to achieve oscillations. We find that the addition of a positive feedback loop on species A in models 2 and 3 increases the ability of the system to achieve limit cycle behavior, but no significant increase in the posterior probability is provided by increasing the feedback strength. This is in conflict with the original study that found that model 3 outperformed model 2 \cite{Tsai:2008p56}. Our approach does sample parameter space predominantly in regions where the desired behavior is more likely, rather than entirely at random as was done in the previous study; on balance this suggests that the posterior probability for delivering robust oscillations is approximately the same for models 2 and 3.
\par
More insight can be gained into this discrepancy by specifying a particular frequency and amplitude of the oscillator as the desired output behavior. Figures \ref{fig:osc1}C and D show the model posterior probability after requiring an amplitude of 0.1 and a frequency of 1.0 Hz on species A and C respectively. The first thing to note is that the model posterior is significantly different in the two cases. When the constraints are applied to species A, model 3 is favored with the increase in feedback strength {\em decreasing} the ability to reach the specified behavior. When the conditions are applied to species C (and species B by symmetry) we get a posterior that more resembles the original findings; that the increase in feedback strength does indeed increase the ability to reach the specified oscillations. Thus the posterior for the Hopf bifurcation behavior represents a sum over all possible oscillator characteristics; in a manner that is reminiscent of Bayesian model averaging.
\par
If we examine the posterior distribution and the principal component analysis for model 2 to achieve Hopf bifurcations (Figures S4 and S5), we can see that the parameters $k_{1}$ and $k_{3}$, which are the strengths of the deactivating reactions on nodes A and B, are constrained to be similar in magnitude to $k_{5}$. We also see that within this model the feedback strength, $k_{7}$, does not affect the dynamics significantly. Here, and elsewhere, we can use the posterior distributions in order to gain insights into the sensitivity and robustness of the system to variations in parameters, irrespective of whether the system's dynamics are deterministic or stochastic: our ABC SMC framework allows us to extract such information on the fly as part of the sequential design process. 
\section{Bacterial two component systems}
Two component systems (TCS) allow bacteria to sense external environmental stimuli and relay information into the cell, e.g. to the gene expression apparatus. They consist of a histidine kinase (HK) that autophosphorylates upon interaction with a specific stimulus. The phosphate group is then passed on to a cognate response regulator protein (RR)  which, once phosphorylated, can regulate transcription  \cite{Stock:2000p2218}.
\par
\begin{figure*}[ht]
\begin{center}
\centerline{ \includegraphics[width=1.0\textwidth]{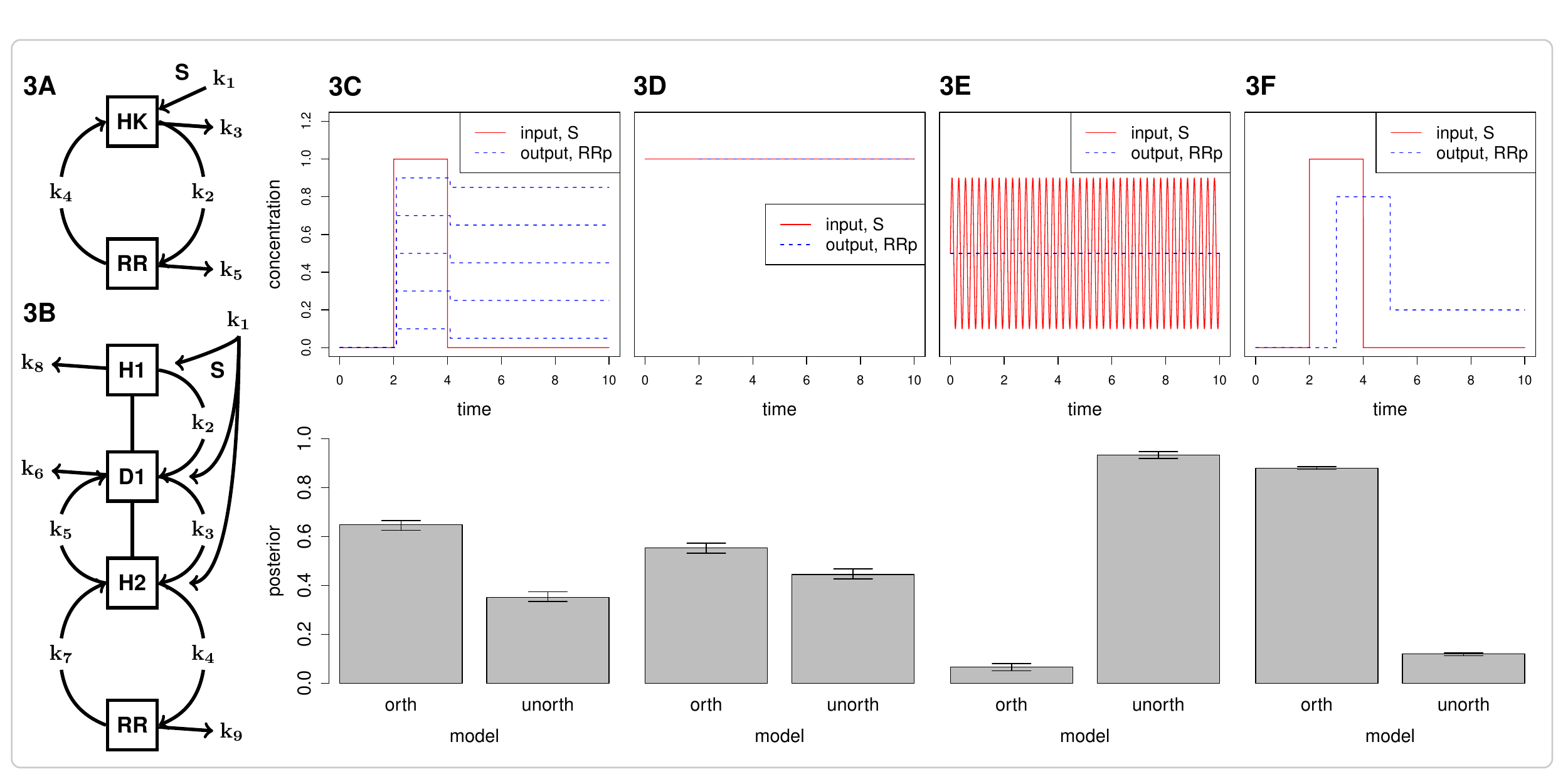} }
\caption{Bacterial two component systems. A) Orthodox system where $HK$ denotes the histidine kinase and $RR$ the response regulator, both of which have phosphorylated forms, $HKp$ and $RRp$. Arrows represent reactions involving phosphate groups and the $k_{i}$ represent the rate parameters. $S$ is the input stimulus signal that causes autophosphorylation of the histidine kinase. B) Phosphorelay system with three phosphate-binding domains, where $H$, $D$ refer to histidine and aspartate domains respectively. C-F) Specified input-output behavior (above) and posterior probabilities for the two designs to achieve it (below). The input signal corresponds to the stimulus, $S$, and the output signal is represented by the concentration of phosphorylated response regulator, $RRp$. The error bars indicate the variability in the marginal model posteriors over three separate runs.}
\label{fig:tcs1}
\end{center}
\end{figure*}

Naturally occurring TCS differ in the number of phosphate binding domains. In the most common form there are two phosphate binding domains (Figure \ref{fig:tcs1}A) but an alternative form exists that consists of a phosphorelay mechanism with four phosphate binding domains (Figure \ref{fig:tcs1}B). These shall be referred to as the orthodox and unorthodox TCSs, respectively \cite{Kim:2006p2084}. The reason for the existence of two forms of TCS remains largely unknown but it has been demonstrated that the phosphorelay is robust to noise and can provide an ultrasensitive response to stimuli \cite{Kim:2006p2084,CsikaszNagy:2010p4748}, whereas the orthodox system can provide behavior that is independent 
of the concentrations of its components \cite{Shinar:2007p2223}.  Here we have applied the Bayesian approach to directly compare the ability of  orthodox and unorthodox designs to achieve various input-output behaviors, using ODE models similar to ones described previously \cite{Kim:2006p2084} (see appendix D). 

Figures \ref{fig:tcs1}C-F show four types of behaviors that may be desired in synthetic TCS systems (e.g. for bioremediation or biopharmaceutical applications), and the corresponding posterior probabilities of the orthodox and unorthodox models to achieve them. In Figure \ref{fig:tcs1}C the specified behavior is that of a fast response to a square pulse input signal. That is the output should show a maximum within 0.1 seconds after the pulse starts and a minimum 0.1 seconds after the pulse ends. As can be seen from the posterior probabilities, both models achieve this behavior easily, as one would expect from a signaling system, with the orthodox system slightly outperforming the unorthodox system. In Figure \ref{fig:tcs1}D the ability of the two systems to achieve a steady output state for $t>2$ seconds under a constant input signal is examined, and again both systems perform comparably with the orthodox system appearing slightly more favorable. 

In Figure \ref{fig:tcs1}E the input signal is a high frequency sinusoid with a mean of 0.6, and the desired output is a constant signal with the same mean and root mean square $< 0.3$; this would mimic a system that is robust to high-frequency noise. The output trajectories at some intermediate and final populations are shown in Figure S6. In this example the unorthodox system clearly outperforms the orthodox system, which indicates the increased robust to noisy signals that comes with the relay architecture \cite{Kim:2006p2084}. But the direct comparison of the two models' ability to cope with noise, which is becoming possible in this approach, also reveals some unexpected characteristics:  inspection of the posterior distribution (Figure S7) shows that all the dephosphorylation reaction rates, $k_6,k_8,k_9$, are minimized while the rate of the signal induced autophosphorylation ($k_1$) is large. Thus the noise reduction mechanism in the unorthodox system works by saturating the system. 

In Figure \ref{fig:tcs1}F the input is again a step function but the output is more specific; it much reach to $> 0.8$ and drop to $<0.2$ within 0.5 seconds of the pulse start and end, respectively, thus approximately reproducing the input. Figure S8 shows the evolution of the system in this case. Here the orthodox model clearly outperforms the unorthodox model. Inspection of the posterior distribution (Figure S9) shows that both the rate of the signal induced autophosphorylation and the rate of phosphorylation of the response regulator by the histidine kinase, $k_1,k_2$, are large while the rate of dephosphorylation of the response regulator, $k_5$, is small. This ensures that the shape of the signal is transferred faithfully through the system.

\section{Stochastic genetic toggle switch without cooperativity}
\begin{figure*}[ht]
\begin{center}
\centerline{ \includegraphics[width=\textwidth]{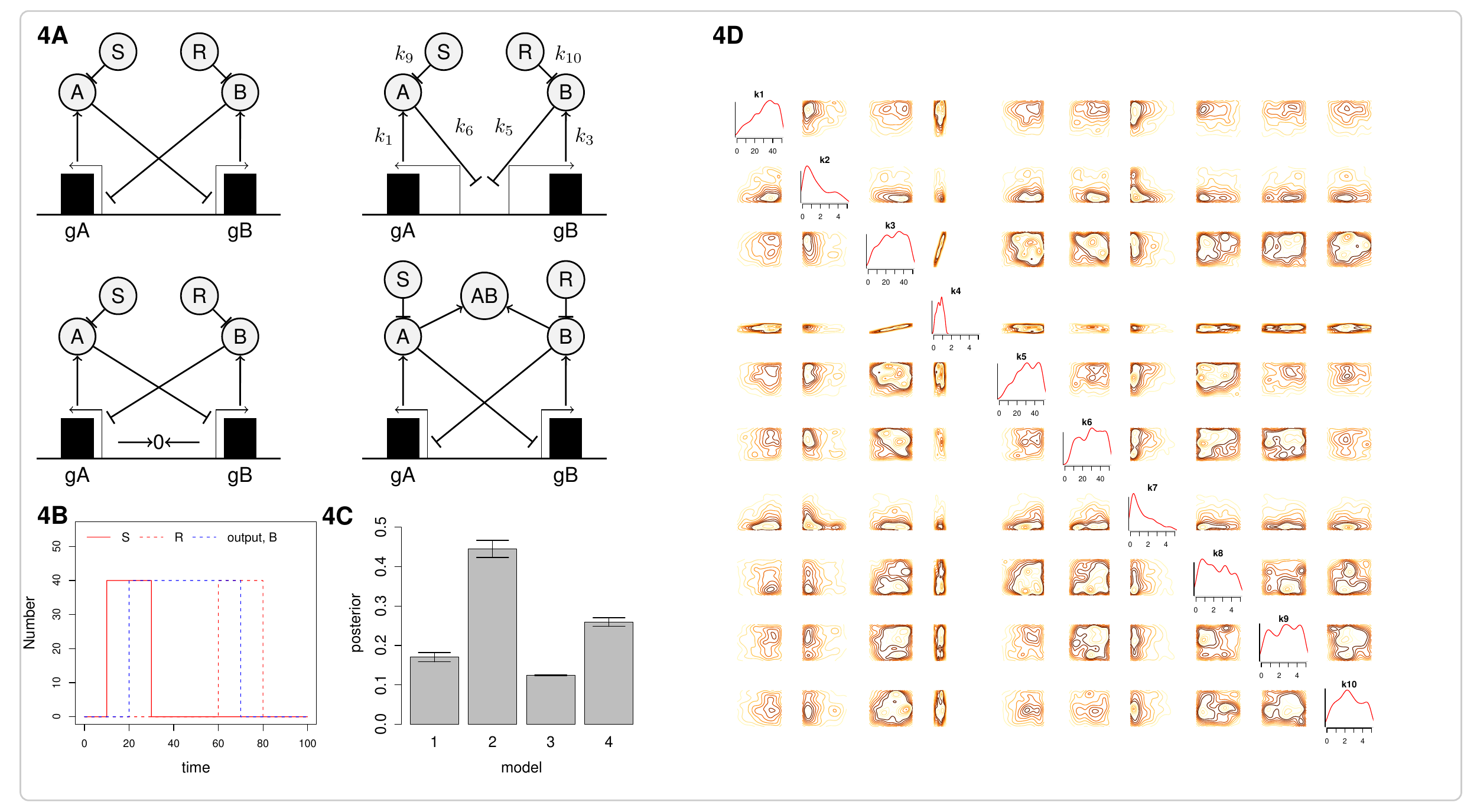} }
\caption{Stochastic toggle switch. A) Four different designs for a toggle switch without cooperative binding. Going clockwise from top left we have the basic switch, the exclusive switch where there is only one repressor site, the basic switch with bound repressor degradation (BRD), and the basic switch with a protein-protein interaction. Genes $gA,gB$ express proteins $A,B$ and $S,R$ represent inducer molecules. The species $AB$ represents complex formation. The rate constants for the reactions are shown for the exclusive switch (protein degradation and transcription factor dissociation from the promoter are not shown). B) The specified input-output behavior. C) The posterior probabilities for each model to achieve the toggle switch behavior. D) Parameter posterior distribution, represented by univariate and bivariate marginal distributions, for model 2 (exclusive switch). }
\label{fig:switch1}
\end{center}
\end{figure*}

The genetic toggle switch is a synthetic realization of a bistable switch that forms the basis of cellular memory \cite{Gardner:2000p2697}. It is formed by two genes $A$ and $B$, whose respective proteins repress the production of the other protein; protein $A$ represses the production of protein $B$ and {\em vice versa} (Figure \ref{fig:switch1}). The presence of an interaction with inducer molecules allows the system to switch between steady states with the probability of spontaneous switching low enough such that, in the absence of an interaction, the system will effectively remain in its appropriate steady state indefinitely. 
\par
Here we consider four versions of the stochastic genetic toggle switch that are all bistable without the requirement for cooperative binding of the proteins to the gene promoter \cite{Lipshtat:2006p5240}. Note that the deterministic models are not necessarily bistable; these are shown in Figure \ref{fig:switch1}A and consist of the basic toggle switch, an exclusive version containing only one promoter, a version that includes bound repressor degradation (BRD), and a version containing a protein-protein interaction between $A$ and $B$ with the resulting complex nonfunctional. The additional reactions are are always to reduce the probability of the 'deadlock' state where both $A$ and $B$ are bound to the promotors of $B$ and $A$ respectively \cite{Lipshtat:2006p5240}. We modeled the switches using a continuous time Markov jump process which obeys the chemical master equation. Only protein level reactions were modeled which makes the models simpler (and faster to simulate) while retaining the important behavior. The stochastic models for all four switches are given in appendix E. For such complicated stochastic dynamical systems the advantages of a Bayesian perspective over conventional model design strategies (based e.g. on optimization) come to the fore: without an appreciation of the whole distribution choice of the best model would be subject to considerable uncertainty. 

Figure \ref{fig:switch1}B shows the desired toggle switch behavior. The inducer $S$ is added between $t=10$ and $t=30$, after which the level of protein $B$ should reach a steady state with a mean number of 40. Between $t=60$ and $t=80$ the inducer $R$ is added after which the level of protein $B$ should drop to zero. The inducer numbers are both assumed to be 40 molecules which is fixed in the specified range. The desired output behavior was specified via the two component distance metric, defined to be $\epsilon = \{d_1,d_2 \}$,
$$ d_1 =  \sqrt{ \frac{ \sum_{t \in \alpha} (x_{t} - y )^2 }{ n_{\alpha} } } \hskip5mm d_2 =  \sqrt{ \frac{ \sum_{t \in \beta} (x_{t} - 0 )^2 }{ n_{\beta} } },
$$ 
where $x_{t}$ is the number of protein $B$ at time $t$, $y$ is the target (here fixed at 40), $\alpha = \{t :  30 < t \le 60 \}$, $\beta = \{t :  0 < t \le 10 \; \text{and} \; 80 < t \le 100 \}$, $n_{\alpha} = \#\alpha$ and $n_{\beta} = \#\beta$. The final population was defined to be $\epsilon = \{7.0,0.05\}$. Here $d_{1}$ represents the distance in the "on" region and $d_{2}$ represents the distance in the ``off" region.

Figure S10 shows the evolution of the stochastic simulations towards the desired behavior. Figure \ref{fig:switch1}C shows the posterior probabilities for each system to achieve the specified behavior, and in particular demonstrates that the exclusive toggle switch outperforms all the others. This chimes with intuition, since the exclusive switch removes the possibility of the deadlock state without the addition of any extra reactions. The fact that the BRD switch performs worse than the original toggle shows that the addition of the two extra degradation reactions does not offer a great enough performance increase for the extra parameters, a manifestation of the parsimony principle or Occam's razor which is inherent to the Bayesian model selection framework used here. 
\par
The reactions that comprise the exclusive switch are given by
\begin{align*}
gA & \xrightarrow{k_1} gA + A & A  & \xrightarrow{k_2} \O{} \\
gB & \xrightarrow{k_3} gB + B & B & \xrightarrow{k_4} \O{} \\
A + gB + P& \xrightarrow{k_5} A-gB & B + gA + P& \xrightarrow{k_6} B-gA \\
A-gB & \xrightarrow{k_7} A + gB + P & B-gA & \xrightarrow{k_8} B + gA + P\\
S+A & \xrightarrow{k_9} S-A & R + B & \xrightarrow{k_{10}} R-B,
\end{align*}
where $gA, gB$ represent the gene promotors for protein $A,B$ respectively and are fixed at one copy, $A-gB,B-gA$ represent the bound transcription factors and $S,R$ represent the inducer molecules. The $P$ species, fixed to be one copy, ensures only $A$ or $B$ can be bound at any one time. Examination of the posterior distribution for this model, Figure \ref{fig:switch1}D, clearly shows a large correlation between $k_3$ and $k_4$, which are the production and decay rates of protein $B$, respectively. This is clearly seen in the principal components (Figure S11) through the combination $k_3+k_4$ dominating the first PC and the combination $k_4-k_3$ dominating the last PC. Thus the system is sensitive to only the difference in these rates which is typical in birth-death processes.

\section{Discussion}
In this paper we have presented a new method for the design of synthetic biological systems employing ideas from Bayesian statistics. We have demonstrated its utility and generality on three different systems spanning biochemical, signaling and genetic networks, as well as oscillatory systems.  This method has advantages over traditional design approaches in that the modeling is incorporated directly into the design stage. The statistical nature of the method has many attractive features including the handling of stochastic systems, the ability to perform model selection and the handling of parameter uncertainty in a well defined manner. We used the ABC-SysBio software \cite{Liepe:2010p2858} which takes as input a set of SBML files and as such can be used by bioengineers and experimentalists to rationally compare their competing designs for a system. By using this method we hope that the implementation time of synthetic systems can be reduced by defining a program of experimental work based on the posterior probabilities of each design. 
\par
Monte Carlo sampling of parameter spaces has been used to assess the robustness of engineered and biological systems in the past. But like in the statistical case, simple Monte Carlo sampling tends to waste too much effort and time on those regions which are of no real interest for reverse-engineering or design purposes. Our statistically based sequential approach homes in onto those regions where the probability of observing the desired behavior is appreciable. This allows us a more nuanced comparative assessment of different design proposals, especially when dynamics are expected (or indeed desired) to exhibit elements of stochasticity. And the Bayesian model selection approach automatically strikes a balance between the systems' abilities to generate the desired behavior effectively but also robustly. 
\par
Further developments will include the incorporation of methods for model abstraction to reduce computation time \cite{Myers:2009p2142} and to handle a database of standard parts as in other existing design software systems \cite{Hill:2008p3637, Marchisio:2008p2206}. Moreover, it is also possible to include the generation of novel structures (by e.g. using stochastic-context free grammars \cite{baldi:1} to propose alterations to a reaction/interaction network) as part of the design process. Just like in the case where ideas from control engineering and statistics can gainfully be combined in order to reverse-engineer the structure of naturally evolved biological systems, we feel that in the design of synthetic systems such a union will also be fruitful. 

\appendix
\section{Approximate Bayesian Computation : ABC SMC}
Here we outline the background behind approximate Bayesian computation (ABC) and describe the ABC SMC algorithm \cite{Toni:2009p606}, which is implemented in the software package ABC-SysBio \cite{Liepe:2010p2858}.  ABC methods have been developed to infer posterior distributions in cases where likelihood functions are computationally intractable or too costly to evaluate. They replace the calculation of the likelihood with a comparison between observed and simulated data. 

\subsection{Background}
Let $\theta \in \Theta$ be a parameter vector with prior $\pi(\theta)$ and $f(y|\theta)$ be the likelihood of the data $y \in \mathcal{D}$. In Bayesian inference we are interested in the posterior density
\begin{equation*}
\pi(\theta|y) = \frac{  f(y|\theta) \pi(\theta) }{ \int_{\Theta} f(y|\theta) \pi(\theta) d\theta} .
\end{equation*}
Now imagine the case where we cannot write down the likelihood in closed form but we can simulate from the data generating model. We can proceed by first sampling a parameter vector from the prior, $\theta^* \sim \pi(\theta)$, and then sampling a data vector, $x^*$, from the model conditional on $\theta^*$, ie $x^*  \sim f(x|\theta^*)$. This alone gives the joint density $\pi(\theta,x)$. To obtain samples from the posterior distribution we must condition on the data $y$ and this is done via an indicator function, i.e.
\begin{equation*}
\pi(\theta, x |y) = \frac{ \pi(\theta) f(x|\theta) \mathbb{I}_{\mathcal{A}_y}(x) } {\int_{\mathcal{A}_y \times \Theta} \pi(\theta) f(x|\theta) dx d\theta},
\end{equation*}
where $\mathbb{I}_{\mathcal{B}}(z)$ denotes the indicator function and is equal to 1 for $z \in \mathcal{B}$. Here $\mathcal{A}_{y} = \{x \in \mathcal{D}: x = y \} $, so the indicator is equal to one when the simulated data and the observed data are identical. This forms a rejection algorithm, and in this instance the accepted $\theta^{*}$ are from the true posterior density $\pi(\theta|y)$.

For most models it is impossible to achieve simulations with outputs in the subset $\mathcal{A}_{y}$ and so an approximation must be made. This is the basis for ABC. In the first instance we can replace $\mathcal{A}_{y}$ by $\mathcal{A}_{y,\epsilon} = \{x \in \mathcal{D}: \rho(x,y) \le \epsilon\}$ where $\rho : \mathcal{D}\times \mathcal{D} \rightarrow \mathbb{R}^{+}$ is a distance function comparing the simulated data to the observed data. We then have
\begin{equation*}
\pi_{\epsilon}(\theta, x |y) = \frac{ \pi(\theta) f(x|\theta) \mathbb{I}_{\mathcal{A}_{y,\epsilon}}(x) } {\int_{\mathcal{A}_{y,\epsilon} \times \Theta} \pi(\theta) f(x|\theta) dx d\theta},
\end{equation*}
where $\pi_{\epsilon}$ is an approximation to the true posterior distribution. The rationale behind ABC is that if $\epsilon$ is small then the resulting approximate posterior, $\pi_{\epsilon}$, is close to the true posterior. Often, for complex models or stochastic systems, the subset $\mathcal{A}_{y,\epsilon}$ is still too restrictive. In these cases we can resort to comparisons of summary statistics. We now specify the subset $\mathcal{A}_{y,\eta,\epsilon} = \{x \in \mathcal{D}: \rho_{S}(x,y) \le \epsilon\}$ where $\eta : \mathcal{D} \rightarrow \mathcal{S}$ is a summary statistic and the distance function now takes the form $\rho_{S} : \mathcal{S}\times \mathcal{S} \rightarrow \mathbb{R}^{+}$. We often write the marginal posterior distribution as $\pi(\theta| \rho(x^{*},y) \le \epsilon)$.

\subsection{ABC SMC}
The simplest ABC algorithm is known as the ABC rejection algorithm \cite{Pritchard:1999p5562} and proceeds as follows\\\\
\begin{tabular}{rl}
R1& Sample $\theta^*$ from $\pi(\theta)$. \\
R2 & Simulate a dataset $x^*$ from $f(x|\theta^*)$. \\
R3 & If $\rho (x^*,y) \le \epsilon$ accept $\theta^*$, otherwise reject. \\
R4 & Return to R1. \\
\end{tabular}\\\\
This gives draws from $\pi_{\epsilon}$ but can be very inefficient in high dimensional models or when the overlap between the prior and posterior distributions is small. One way to improve the efficiency of the rejection algorithm is to perform sequential importance sampling (SIS) \cite{DelMoral:2006p1563}. In SIS, instead of sampling directly from the posterior distribution, sampling proceeds via a series of intermediate distributions. The importance distribution at each stage is constructed from a perturbed version of the previous population. This approach can be used in ABC and the resultant algorithm is known as ABC SMC \cite{Toni:2009p606}. Described here is a slightly modified version that automatically calculates the $\epsilon$ schedule and as such, only the final value, $\epsilon_{T}$, needs be specified. To obtain $N$ samples $\{ \theta^1, \theta^2, \theta^3 ...., \theta^N \}$ (known as particles) from the posterior, defined as, $\pi(\theta| \rho(x^{*},y) \le \epsilon_{T})$, proceed as follows\\\\
\begin{tabular}{rl}
 S1 & Initialize $\epsilon = \infty$\\
      & Set the population indicator $t=0$ \\
S2.0 & Set the particle indicator $i=1$\\
S2.1 &If $t=0$, sample $\theta^{**}$ independently from $\pi(\theta)$ \\
         &If $t>0$, sample $\theta^*$ from the previous population $\{ \theta_{t-1}^{i} \}$ with weights $w_{t-1}$.\\
         & Perturb the particle, $\theta^{**} \sim K_t(\theta|\theta^{*})$ where $K_t$ is the perturbation kernel.\\
         &If $\pi(\theta^{**}) = 0$, return to S2.1\\
         &Simulate a candidate dataset $x^{*} \sim f(x|\theta^{**})$. \\
         &If $\rho(x^*,y) > \epsilon$ return to S2.1\\
S2.2 & Set $\theta^{i}_{t} = \theta^{**}$ and $d^{i}_t = \rho(x^*,y)$, calculate the weight as \\
	& 
$
w_{t}^{i} = \left\{ 
\begin{array}{rl}
 1 \qquad \qquad &\mbox{ if $t=0$} \\  
  \frac{ \pi(\theta_{t}^{i})}{ \sum^{N}_{j=1} w_{t-1}^{j} K_{t}(\theta_{t}^{i} | \theta^{j}_{t-1} ) } &\mbox{ if $t>0$}
\end{array} \right.
$ \\\\
      & If $i<N$, set $i=i+1$, go to S2.1\\
S3 & Normalize the weights. \\
      & Determine $\epsilon$ such that $Pr(d_t \le \epsilon ) =0.9$. \\
      & If $\epsilon > \epsilon_{T}$, set $t=t+1$, go to S2.0.
\end{tabular}\\\\
\noindent Here $K_{t}(\theta|\theta^{*})$ is the component-wise random walk perturbation kernel that, in this study, takes the form $K_{t}(\theta^{*}|\theta) = \theta + U(-\delta,\delta)$ where $\delta = \frac{1}{2}\text{range} \{\theta_{t-1}\}$. The denominator in the weight calculation can be seen as the probability of observing the current particle given the previous population.

\subsection{Model selection}
In Bayesian inference comparison of a discrete set of models can be be performed using the marginal posterior. Consider the joint space defined by $(M,\theta) \in \mathcal{M} \times \Theta_{\mathcal{M}} $; Bayes theorem can then be written
\begin{align*}
\pi(M | y) = \frac{ f(y | M)  \pi(M)}{ \int_{\mathcal{M}} f(y | M')  \pi(M')dM' } = \frac{ f(y | M)  \pi(M)}{ \sum_{\mathcal{M}} f(y | M')  \pi(M') },
\end{align*}
where $f(y | M)$, the marginal likelihood, can be written
\begin{align*}
f(y | M) = \int_{\Theta_{\mathcal{M}} } \pi(\theta | M ) f(y | \theta, M) d\theta.
\end{align*}
\noindent  Therefore the posterior probability of a model is given by the normalized marginal likelihood which may or may not be weighted depending on whether the prior over models is informative or uniform respectively. It has recently been noted that model selection using summary statistics can be problematic because the summary statistic must be sufficient for the joint space, $\{M, \theta\}$, rather than just $\theta$ \cite{Robert:2011p7915}. This is not a concern here since in all our examples we use the full data set with no summary or we {\it define} our posterior distributions through the summary statistics. 

Model selection can be incorporated into the ABC framework by introducing the model indicator $M$ and proceeding with inference on the joint space. For example, the ABC rejection algorithm with model selection \cite{Grelaud:2009p5626} proceeds as follows\\\\
\begin{tabular}{rl}
MR1& Sample $M^*$ from $\pi(M)$. \\
MR2 & Sample $\theta^*$ from $\pi(\theta| M^*)$. \\
MR3 & Simulate a dataset $x^*$ from $f(x|\theta^*, M^*)$. \\
MR4 & If $\rho (x^*,y) \le \epsilon$ accept $(M^*,\theta^*)$, otherwise reject. \\
MR5 & Return to R1. \\
\end{tabular}\\\\
Once $N$ samples have been accepted an approximation to the marginal posterior, $\pi(M=m|y)$, is given by
\begin{align*}
\pi(M=m|y) = \frac{ \text{ \#accepted m }}{ N}.
\end{align*}
Model selection can also be incorporated into the ABC SMC algorithm \cite{Toni:2010p2027}. To obtain $N$ samples \\$\{ (M, \theta)^1, (M,\theta)^2, (M,\theta)^3 ...., (M,\theta)^N \}$ from the posterior, defined as, $\pi(M,\theta| \rho(x^{*},y) \le \epsilon_{T})$, proceed as follows\\\\
\begin{tabular}{rl}
 MS1 & Initialize $\epsilon = \infty$\\
      & Set the population indicator $t=0$ \\
MS2.0 & Set the particle indicator $i=1$\\
MS2.1 &If $t=0$, sample $(M^{**},\theta^{**})$ from the prior $\pi(M,\theta) = \pi(M)\pi(\theta|M)$. \\
         &If $t>0$, sample $M^*$ with probability $P_{t-1}(M^*)$ and perturb $M^{**} \sim KM_{t}(M|M^{*})$.\\
         &Sample $\theta^*$ from the previous population $\{ \theta(M^{**})_{t-1} \}$ with weights $w_{t-1}$. \\
         &Perturb the particle, $\theta^{**} \sim K_{t,M^{**}}(\theta|\theta^{*})$ where $K_{t,M}$ is the perturbation kernel.\\
         &If $\pi(M^{**},\theta^{**}) = 0$, return to MS2.1\\
         &Simulate a candidate dataset $x^{*} \sim f(x|M^{**},\theta^{**})$. \\
         &If $\rho(x^*,y) > \epsilon$ return to MS2.1\\
MS2.2 & Set $(M,\theta)^{i}_{t} = (M^{**},\theta^{**})$ and $d^{i}_t = \rho(x^*,y)$, calculate the weight as \\
	& 
$
w_{t}^{i}(M^{i}_{t},\theta^{i}_{t}) = \left\{ 
\begin{array}{rl}
 1 \qquad &\mbox{ if $t=0$} \\
  \frac{ \pi(M^{i}_{t},\theta_{t}^{i})}{S_1 S_2} &\mbox{ if $t>0$}
\end{array} \right.
$ \\
& where \\
& $
S_1 = \sum_{j \in \mathcal{M}} P_{t-1}(M^{j}_{t-1})KM_{t} (M_{t}^{i} | M_{t-1}^{j} )
$\\
& and \\
& \multirow{2}{*}{ $
S_2 = \sum_{k \in M^{i}_t = M_{t-1}} \frac{ w_{t-1}^{k} K_{t,M^{i}}(\theta^{i}_{t}|\theta_{t-1}^{k}) }{ P_{t-1}(M^{i}_t = M_{t-1}) }
$} \\\\

      & If $i<N$, set $i=i+1$, go to MS2.1\\
S3 & Normalize the weights. \\
      & Obtain the marginal model probabilities given by\\\\
      &
$
P_t(M_t=m) = \sum_{k \in M^{i}_t = M_{t-1}} w_{t}^i(M_{t}^i,\theta_t^i)
$\\\\
      & Determine $\epsilon$ such that $Pr(d_t \le \epsilon ) =0.9$. \\
      & If $\epsilon > \epsilon_{T}$, set $t=t+1$, go to MS2.0.
\end{tabular}\\\\

\noindent There are two obvious additions to the algorithm when compared to parameter inference. The model kernel, $KM_{t}$, perturbs the resampled models using a multinomial distribution, and the additional term in the weight denominator accounts for the probability of observing the current model given the previous population.

\subsection{Prior distribution}
The prior distribution encodes our knowledge of the system and should be set according to known biochemical properties. However, often the kinetic parameters are not well known and can be very difficult or even impossible to measure {\em in vivo}. In these cases we make the prior distribution non informative by specifying a large range over possible, biophysically and biochemically plausible values. As more information becomes available, through experimental studies or otherwise, the prior can be updated to reflect our increased knowledge of the system. Interestingly, for some systems, our design method could help to constrain kinematic parameters where experimental data are unavailable.

\subsection{The distance function and output tolerance}
In system design we would rarely insist on achieving the true posterior distribution corresponding to $\epsilon=0$, but would like to reach the objective within some tolerance. A theorem due to Wilkinson (2008) \cite{Wilkinson:2008p3789} states that if we assume that the data can be considered as 
\begin{align*}
y = \eta( \hat \theta)  + e,
\end{align*}
where $\eta(\hat \theta)$ is a draw from the model at the 'best' input and $e$ is an additive, independent error, then the approximate posterior distribution, $\pi(\theta| \rho(x^{*},y) \le \epsilon)$ can be interpreted as the 'true' posterior $\pi(\hat \theta|y)$. While the independence assumption is not always true, this theorem provides some insight into the relationship between the final $\epsilon$ value and the tolerance on our specified behavior. For example when using uniform kernels, as in this study, if our desired output behavior is a constant of 0.5 and we finish the inference at $\epsilon =0.05$ our final trajectories will be distributed $U(0.45,0.55)$ giving a tolerance of $\pm 10\%$ on the output behavior. This can be used when considering our desired output objectives. To achieve other error distributions, such as Gaussian errors, we can always explicitly specify the error model in the design objectives.

\subsection{Deterministic models}
Inference for deterministic models such as ordinary differential equations can be problematic since there is a one to one relationship between the parameter vector $\theta$ and the data set $x$. Therefore, in the absence of observational error, the posterior distribution resembles a delta function, $\delta(\theta - \hat \theta)$ where $\hat x = f(\hat \theta)$ is data 'closest' to $y$. An additional problem for ABC methods is that the minimum distance, $\rho(\hat x,y)$, is greater than zero \cite{Toni:2010thesis}. However, in practice, observational data have associated experimental errors and when this is included explicitly in the model, the problem is resolved. In the case of systems design, we omit the explicit error model for clarity, but note that it could be included with assumptions on the form of the distribution.

\section{Biochemical adaptation}
\subsection{Models}
We used the same models as those used in \cite{Ma:2009p2043}, which are enzymatic reactions assuming Michaelis-Menten kinetics. Below we give the full models including cooperativity but the more specific case of no cooperativity is when the exponents, $n_i$, are set to one. Here $A, B, C $ denote the concentrations of the active form of the species and $(1-A), (1-B), (1-C)$ the concentrations of the inactive form. Species $E_i$ and $F_i$ refer to background activating and deactivating enzymes respectively and are assumed to have a constant concentration of 0.5. The models were simulated in the range $0 \le t \le 200$. 
\\ {\bf Design 1}
\begin{align*}
\frac{dA}{dt} &=  I k_{IA} \frac{(1-A)^{n_{IA}} }{(1-A)^{n_{IA}} + K_{IA}^{n_{IA}} } - F_{A} k_{FA}\frac{ A^{n_{FA}}}{A^{n_{FA}}+ K_{FA}^{n_{FA}} }\\
\frac{dB}{dt} & = C k_{CB} \frac{(1-B)^{n_{CB}} }{(1-B)^{n_{CB}} + K_{CB}^{n_{CB}}} - F_{B} k_{FB}\frac{B^{n_{FB}}}{B^{n_{FB}} + K_{FB}^{n_{FB}} }\\
\frac{dC}{dt} & = A k_{AC} \frac{(1-C)^{n_{AC}} }{(1-C)^{n_{AC}} + K_{AC}^{n_{AC}} } - B k_{BC}\frac{C^{n_{BC}} }{C^{n_{BC}}+ K_{BC}^{n_{BC}} }
\end{align*}

\noindent {\bf Design 2}
\begin{align*}
\frac{dA}{dt} &=  I k_{IA} \frac{(1-A)^{n_{IA}} }{(1-A)^{n_{IA}} + K_{IA}^{n_{IA}} } - F_{A} k_{FA}\frac{ A^{n_{FA}}}{A^{n_{FA}}+ K_{FA}^{n_{FA}} }\\
\frac{dB}{dt} & = E_B k_{EB} \frac{(1-B)^{n_{EB}} }{(1-B)^{n_{EB}} + K_{EB}^{n_{EB}}} - C k_{CB}\frac{B^{n_{CB}}}{B^{n_{CB}} + K_{CB}^{n_{CB}} }\\
\frac{dC}{dt} & = A k_{AC} \frac{(1-C)^{n_{AC}} }{(1-C)^{n_{AC}} + K_{AC}^{n_{AC}} } - B k_{BC}\frac{C^{n_{BC}} }{C^{n_{BC}}+ K_{BC}^{n_{BC}} } - F_{C} k_{FC}\frac{ C^{n_{FC}}}{C^{n_{FC}}+ K_{FC}^{n_{FC}} } 
\end{align*}

\noindent {\bf Design 3}
\begin{align*}
\frac{dA}{dt} &=  I k_{IA} \frac{(1-A)^{n_{IA}} }{(1-A)^{n_{IA}} + K_{IA}^{n_{IA}} } - F_{A} k_{FA}\frac{ A^{n_{FA}}}{A^{n_{FA}}+ K_{FA}^{n_{FA}} }\\
\frac{dB}{dt} & = E_B k_{EB} \frac{(1-B)^{n_{EB}} }{(1-B)^{n_{EB}} + K_{EB}^{n_{EB}}} - C k_{CB}\frac{B^{n_{CB}}}{B^{n_{CB}} + K_{CB}^{n_{CB}} }\\
\frac{dC}{dt} & = B k_{BC}\frac{(1-C)^{n_{BC}} }{(1-C)^{n_{BC}}+ K_{BC}^{n_{BC}} } - A k_{AC} \frac{C^{n_{AC}} }{C^{n_{AC}} + K_{AC}^{n_{AC}} }
\end{align*}

\noindent {\bf Design 4}
\begin{align*}
\frac{dA}{dt} &=  I k_{IA} \frac{(1-A)^{n_{IA}} }{(1-A)^{n_{IA}} + K_{IA}^{n_{IA}} } - B k_{BA}\frac{ A^{n_{BA}}}{A^{n_{BA}}+ K_{BA}^{n_{BA}} }\\
\frac{dB}{dt} & = A k_{AB} \frac{(1-B)^{n_{AB}} }{(1-B)^{n_{AB}} + K_{AB}^{n_{AB}}} - F_B k_{FB}\frac{B^{n_{FB}}}{B^{n_{FB}} + K_{FB}^{n_{FB}} }\\
\frac{dC}{dt} & = A k_{AC}\frac{(1-C)^{n_{AC}} }{(1-C)^{n_{AC}}+ K_{AC}^{n_{AC}} } - F_C k_{FC} \frac{C^{n_{FC}} }{C^{n_{FC}} + K_{FC}^{n_{FC}} }
\end{align*}

\noindent {\bf Design 5}
\begin{align*}
\frac{dA}{dt} &=  I k_{IA} \frac{(1-A)^{n_{IA}} }{(1-A)^{n_{IA}} + K_{IA}^{n_{IA}} } - B k_{BA}\frac{ A^{n_{BA}}}{A^{n_{BA}}+ K_{BA}^{n_{BA}} }\\
\frac{dB}{dt} & = A k_{AB} \frac{(1-B)^{n_{AB}} }{(1-B)^{n_{AB}} + K_{AB}^{n_{AB}}} - F_B k_{FB}\frac{B^{n_{FB}}}{B^{n_{FB}} + K_{FB}^{n_{FB}} }\\
\frac{dC}{dt} & = E_C k_{EC} \frac{(1-C)^{n_{EC}} }{(1-C)^{n_{EC}} + K_{EC}^{n_{EC}} } - A k_{AC}\frac{C^{n_{AC}} }{C^{n_{AC}}+ K_{AC}^{n_{AC}} } 
\end{align*}

\noindent {\bf Design 6}
\begin{align*}
\frac{dA}{dt} &=  I k_{IA} \frac{(1-A)^{n_{IA}} }{(1-A)^{n_{IA}} + K_{IA}^{n_{IA}} } - B k_{BA}\frac{ A^{n_{BA}}}{A^{n_{BA}}+ K_{BA}^{n_{BA}} }\\
\frac{dB}{dt} & = C k_{CB} \frac{(1-B)^{n_{CB}} }{(1-B)^{n_{CB}} + K_{CB}^{n_{CB}}} - F_B k_{FB}\frac{B^{n_{FB}}}{B^{n_{FB}} + K_{FB}^{n_{FB}} }\\
\frac{dC}{dt} & = A k_{AC}\frac{(1-C)^{n_{AC}} }{(1-C)^{n_{AC}}+ K_{AC}^{n_{AC}} } - F_C k_{FC} \frac{C^{n_{FC}} }{C^{n_{FC}} + K_{FC}^{n_{FC}} }
\end{align*}

\noindent {\bf Design 7}
\begin{align*}
\frac{dA}{dt} &=  I k_{IA} \frac{(1-A)^{n_{IA}} }{(1-A)^{n_{IA}} + K_{IA}^{n_{IA}} } - B k_{BA}\frac{ A^{n_{BA}}}{A^{n_{BA}}+ K_{BA}^{n_{BA}} } - F_{A} k_{FA}\frac{ A^{n_{FA}}}{A^{n_{FA}}+ K_{FA}^{n_{FA}} } \\
\frac{dB}{dt} & = E_B k_{EB} \frac{(1-B)^{n_{EB}} }{(1-B)^{n_{EB}} + K_{EB}^{n_{EB}}} - C k_{CB}\frac{B^{n_{CB}}}{B^{n_{CB}} + K_{CB}^{n_{CB}} }\\
\frac{dC}{dt} & = A k_{AC}\frac{(1-C)^{n_{AC}} }{(1-C)^{n_{AC}}+ K_{AC}^{n_{AC}} } - F_C k_{FC} \frac{C^{n_{FC}} }{C^{n_{FC}} + K_{FC}^{n_{FC}} }
\end{align*}

\noindent {\bf Design 8}
\begin{align*}
\frac{dA}{dt} &=  I k_{IA} \frac{(1-A)^{n_{IA}} }{(1-A)^{n_{IA}} + K_{IA}^{n_{IA}} } - B k_{BA}\frac{ A^{n_{BA}}}{A^{n_{BA}}+ K_{BA}^{n_{BA}} } - F_{A} k_{FA}\frac{ A^{n_{FA}}}{A^{n_{FA}}+ K_{FA}^{n_{FA}} } \\
\frac{dB}{dt} & = C k_{CB} \frac{(1-B)^{n_{CB}} }{(1-B)^{n_{CB}} + K_{CB}^{n_{CB}}} - F_{B} k_{FB}\frac{B^{n_{FB}}}{B^{n_{FB}} + K_{FB}^{n_{FB}} }\\
\frac{dC}{dt} & = E_C k_{EC} \frac{(1-C)^{n_{EC}} }{(1-C)^{n_{EC}} + K_{EC}^{n_{EC}} } - A k_{AC}\frac{C^{n_{AC}} }{C^{n_{AC}}+ K_{AC}^{n_{AC}} } 
\end{align*}

\noindent {\bf Design 9}
\begin{align*}
\frac{dA}{dt} &=  I k_{IA} \frac{(1-A)^{n_{IA}} }{(1-A)^{n_{IA}} + K_{IA}^{n_{IA}} } - B k_{BA}\frac{ A^{n_{BA}}}{A^{n_{BA}}+ K_{BA}^{n_{BA}} }\\
\frac{dB}{dt} & = E_B k_{EB} \frac{(1-B)^{n_{EB}} }{(1-B)^{n_{EB}} + K_{EB}^{n_{EB}}} - C k_{CB}\frac{B^{n_{CB}}}{B^{n_{CB}} + K_{CB}^{n_{CB}} }\\
\frac{dC}{dt} & = E_C k_{EC} \frac{(1-C)^{n_{EC}} }{(1-C)^{n_{EC}} + K_{EC}^{n_{EC}}} - A k_{AC}\frac{C^{n_{AC}} }{C^{n_{AC}}+ K_{AC}^{n_{AC}} } 
\end{align*}

\noindent {\bf Design 10}
\begin{align*}
\frac{dA}{dt} &=  I k_{IA} \frac{(1-A)^{n_{IA}} }{(1-A)^{n_{IA}} + K_{IA}^{n_{IA}} } - F_{A} k_{FA}\frac{ A^{n_{FA}}}{A^{n_{FA}}+ K_{FA}^{n_{FA}} }\\
\frac{dB}{dt} & = A k_{AB} \frac{(1-B)^{n_{AB}} }{(1-B)^{n_{AB}} + K_{AB}^{n_{AB}}} - F_B k_{FB}\frac{B^{n_{FB}}}{B^{n_{FB}} + K_{FB}^{n_{FB}} }\\
\frac{dC}{dt} & = B k_{BC}\frac{(1-C)^{n_{BC}} }{(1-C)^{n_{BC}}+ K_{BC}^{n_{BC}} } - A k_{AC} \frac{C^{n_{AC}} }{C^{n_{AC}} + K_{AC}^{n_{AC}} }
\end{align*}

\noindent {\bf Design 11}
\begin{align*}
\frac{dA}{dt} &=  I k_{IA} \frac{(1-A)^{n_{IA}} }{(1-A)^{n_{IA}} + K_{IA}^{n_{IA}} } - F_{A} k_{FA}\frac{ A^{n_{FA}}}{A^{n_{FA}}+ K_{FA}^{n_{FA}} }\\
\frac{dB}{dt} & = A k_{AB} \frac{(1-B)^{n_{AB}} }{(1-B)^{n_{AB}} + K_{AB}^{n_{AB}}} - F_B k_{FB}\frac{B^{n_{FB}}}{B^{n_{FB}} + K_{FB}^{n_{FB}} }\\
\frac{dC}{dt} & = A k_{AC}\frac{(1-C)^{n_{AC}} }{(1-C)^{n_{AC}}+ K_{AC}^{n_{AC}} } - B k_{BC} \frac{C^{n_{BC}} }{C^{n_{BC}} + K_{BC}^{n_{BC}} }
\end{align*}

\subsection{Distance}
The two component distance metric was defined to be $\epsilon = \{E,S^{-1} \}$, where $E$ and $S$ are the adaptation efficiency and sensitivity defined by 
\begin{eqnarray}
E &= & \Bigg\lvert \frac{ (O_2 - O_1)/O_1 }{ (I_2 - I_1)/I_1}  \Bigg\rvert \nonumber \\
S &= & \Bigg\lvert \frac{ (O_{peak} - O_1)/O_1 }{ (I_2 - I_1)/I_1} \Bigg\lvert \nonumber,
\end{eqnarray}
where $I_1, I_2$ are the input values (here fixed at 0.5 and 0.6 respectively), $O_{1}, O_{2}$ are the output steady state levels before and after the input change and and $O_{peak}$ is the maximal transient output level. The final population was defined to be $\epsilon = \{0.1,1.0\}$. 
\subsection{Priors}
The priors on the Michaelis-Menten rates were chosen to correspond to the parameter ranges used in the original study; $ \log k \sim U(-1,1)$ and $\log K \sim U(-3,2)$ \cite{Ma:2009p2043}.

\section{Robust oscillator design}
\subsection{Models}
We used the same models as those used in \cite{Tsai:2008p56}, simulated in the range $0 \le t \le 10$. Again $A,B,C$ denote the concentrations of the active form of the species and $(1-A),(1-B),(1-C)$ the concentrations of the inactive form. The feedback is modeled using Michaelis-Menten kinetics but the conversion of inactive form into active form is assumed to have a constant rate.
\\ {\bf Design 1}
\begin{align*}
\frac{dA}{dt} &=  k_{1}(1-A) - \frac{k_{2}C^{n_1}}{K_{1}^{n_1} + C^{n_1} } A\\
\frac{dB}{dt} &=  k_{3}(1-B) - \frac{k_{4}A^{n_2}}{K_{2}^{n_2} + A^{n_2} } B \\
\frac{dC}{dt} &= k_{5}(1-C) - \frac{k_{6}B^{n_3}}{K_{3}^{n_3} + B^{n_3} } C
\end{align*}
{\bf Designs 2 and 3}
\begin{align*}
\frac{dA}{dt} &= k_{1}(1-A) - \frac{k_{2}C^{n_1}}{K_{1}^{n_1} + C^{n_1} } A + k_{7}(1-A)\frac{A^{n_4}}{K_{4}^{n_4} + A^{n_4}} \\
\frac{dB}{dt} &= k_{3}(1-B) - \frac{k_{4}A^{n_2}}{K_{2}^{n_2} + A^{n_2} } B \\
\frac{dC}{dt} &= k_{5}(1-C) - \frac{k_{6}B^{n_3}}{K_{3}^{n_3} + B^{n_3} } C
\end{align*}
{\bf Designs 4 and 5}
\begin{align*}
\frac{dA}{dt} &= k_{1}(1-A) - \frac{k_{2}C^{n_1}}{K_{1}^{n_1} + C^{n_1} } A - k_{7}A\frac{A^{n_4}}{K_{4}^{n_4} + A^{n_4}}\\
\frac{dB}{dt} &= k_{3}(1-B) - \frac{k_{4}A^{n_2}}{K_{2}^{n_2} + A^{n_2} } B \\
\frac{dC}{dt} &= k_{5}(1-C) - \frac{k_{6}B^{n_3}}{K_{3}^{n_3} + B^{n_3} } C
\end{align*}

\subsection{Distance}
For the direct Hopf bifurcation detection the distance metric was defined to be 
\begin{equation*}
\epsilon = \frac{ \prod_{i} \Re[\lambda_i] }{ \prod_{i}(1-0.99 \exp(-| \Im [\lambda_i] |)) }
\end{equation*}
where $\lambda_i$ is the $i^{th}$ complex eigenvalue of the linearized system in the steady state. Here $\epsilon=0$ represents the location in parameter space where  a limit cycle emerges through a Hopf bifurcation \cite{Chickarmane:2005p4692}. The final population was at $\epsilon=0.001$. 
\par
To investigate the ability to achieve particular amplitude-frequency values, the distance was defined as $\epsilon = \{d_1,d_2,d_3\}$, where
\begin{eqnarray*}
d_1 &= &\sum_{n}  |x_{t_0 + nT} - x_{t_0 + (n-1)T}| \\
d_2 &= & |f_t - f|\\
d_3 &= & |\max x_{t>t_0} - \min x_{t>t_0} - A_{t}|,\\
\end{eqnarray*}
and $n$ is an integer, $f_t$ is the target frequency, $f$ is the frequency determined from the largest component of the Fourier spectrum, $A_t$ is the target amplitude and $t_0$ is a cut to remove initial transients ($=2$s). The final population was defined to be $\epsilon = \{0.05,0.05,0.05\}$.

\subsection{Priors}
The priors were chosen to correspond to parameter ranges used in the original study; $k_{1} \sim U(0,10)$, $k_{2} \sim U(0,1000)$, $k_{3} \sim U(0,10)$, $k_{4} \sim U(0,1000)$, $k_{6} \sim U(0,1000)$, $k_{7} \sim U(0,100)$, $k^{strong}_{7} \sim U(500,600)$, $n_{i} \sim U(1,4)$ and $K_{i} \sim U(0,4)$ \cite{Tsai:2008p56}.

\section{Bacterial two component systems}
\subsection{Models}
The models we used were based on the ones found in \cite{Kim:2006p2084}. All simulations were performed in the range $0 \le t \le 10$.\\
{\bf Orthodox system}\\
We modeled the following reactions
\begin{align*}
HK + S & \xrightarrow{k_1} HKp + S \\
HKp + RR & \xrightarrow{k_2} HK + RRp\\
HKp & \xrightarrow{k_3} HK \\
HK + RRp & \xrightarrow{k_4} HKp + RR\\
RRp & \xrightarrow{k_5} RR.
\end{align*}
Additionally we assumed that the total concentration of $HK_{tot} = HK + HKp$ and $RR_{tot} = RR + RRp$ were equal to one. This resulted in the following ordinary differential equations
\begin{align*}
\frac{d [HK]}{dt} & = k_2 [HKp] [RR] + k_3 [HKp] - k_4 [HK] [RRp] - k_1 [HK] [S] \\
\frac{d [RRp]}{dt} & = k_2 [HKp] [RR] - k_4 [HK] [RRp] -k_5 [RRp].
\end{align*}

\noindent {\bf Orthodox system}\\
We labelled the occupied states of the phosphorelay as 
\begin{center}
\begin{tabular}{cccc}
 & H1 & D1 & H2 \\
 \hline
$HK_1$ & x & x & x\\
$HK_2$ & o & x & x\\
$HK_3$ & x & o & x\\
$HK_4$ & x & x & o \\
$HK_5$ & o & o & x\\
$HK_6$ & o & x & o\\
$HK_7$ & x & o & o\\
$HK_8$ & o & o & o\\
\end{tabular}
\end{center}
where H1, D1 and H2 are the binding domains on the Histidine Kinase and x, o represent an empty, occupied domain respectively. We modeled the following reactions
\begin{align*}
HK + S & \xrightarrow{k_1} HKp + S
\end{align*}
Again we assumed that the total concentration of $HK_{tot} = \sum HK_{i}$ and $RR_{tot} = RR + RRp$ were equal to one. This resulted in the following ordinary differential equations
\begin{align*}
\frac{dHK_1}{dt} & = k_4 [HK_4] [RR] + k_6 [HK_3] - k_7 [HK_1][RRp] + k_8 [HK_2] - k1 [HK_1] [S] \\
\frac{dHK_2}{dt} & = k_4 [HK_6] [RR] + k_6 [HK_5] - k_7 [HK_2] [RRp] - k_8 [HK_2] + k_1[HK_1][S] -k_2 [HK_2] \\
\frac{dHK_3}{dt} & = -k_3 [HK_3] +k_4 [HK_7][RR] + k_5 [HK_4] - k_6 [HK_3] -k_7[HK_3][RRp] +k_8 [HK_5] -k_1[HK_3][S] +k_2[HK_2] \\
\frac{dHK_4}{dt} & = k_3 [HK_3] -k_4 [HK_4][RR] -k_5[HK_4] + k_6[HK_7] +k_7[HK_1][RRp] +k_8[HK_6] -k_1[HK_4][S] \\
\frac{dHK_5}{dt} & = -k_3[HK_3] +k_4 [HK_8][RR] +k_5[HK_6] -k_6 [HK_5] -k_7 [HK_5][RRp] -k_8 [HK_5] +k_1 [HK_3][S] \\
\frac{dHK_6}{dt} & = k_3 [HK_5] -k_2 [HK_6] -k_4 [HK_6][RR] -k_5 [HK_6] +k_6[HK_8] +k_7 [HK_2][RRp] -k_8 [HK_6] +k_1 [HK_4] [S] \\
\frac{dHK_7}{dt} & = k_2 [HK_6] - k_4 [HK_7][RR] -k_6 [HK_7] + k_7 [HK_3][RRp] +k_8 [HK_8] - k_1 [HK_7][S] \\
\frac{dRRp}{dt} & = k_4 [RR]( [HK_4] + [HK_6] + [HK_7] + [HK_8] ) -k_7 [RRp]( [HK_1] + [HK_2] + [HK_3] + [HK_5] ) -k_9 [RRp]
\end{align*}

\subsection{Distance}
The distance functions for input-output behaviors $\epsilon_{1-4}$ were defined to be
\begin{eqnarray*}
\epsilon_{1} &= & \Bigl\lbrace  \mathbb{H}(0)(\argmax_t x_t - 2.0), \mathbb{H}(0)(\argmin_t x_t - 4.0) \Bigl\rbrace \\
\epsilon_{2} &= & \sqrt{ \sum_{t} (x_{t} - 1.0 )^2 }  \\
\epsilon_{3} &= & \sqrt{ \sum_{t} (x_{t} - 0.5)^2 } \\
\epsilon_{4} &= & \Bigl\lbrace  \epsilon_1, \mathbb{H}(0)(1 - \max x_t -0.2), \mathbb{H}(0)(\min x_t-0.2) \Bigl\rbrace
\end{eqnarray*}
where $\mathbb{H}(0)$ is the Heaviside function ensuring the distance is positive.

\subsection{Priors}
The priors on all variables were distributed as $U(0,1000)$.

\section{Stochastic genetic toggle switch}
\subsection{Models}
We modeled each toggle switch using a continuous time Markov jump process which obeys the chemical master equation. We neglected processes at the RNA level and just modeled at the protein level. This makes the models simpler while retaining all the relevant behavior. In all the following $gA,gB$ represent the gene promotor for protein $A,B$ respectively and are fixed at one copy. $A-gB,B-gA$ represent the bound transcription factors and $S,R$ represent the switch and reset signals. Because the concentration of these are fixed they have the effect of removing $A$ and $B$ from the system respectively. The models were simulated in the range $0 \le t \le 200$\\
{\bf Design 1}
\begin{align*}
gA & \xrightarrow{k_1} gA + A \\
A  & \xrightarrow{k_2} \O{} \\
gB & \xrightarrow{k_3} gB + B \\
B & \xrightarrow{k_4} \O{} \\
A + gB & \xrightarrow{k_5} A-gB \\
B + gA & \xrightarrow{k_6} B-gA \\
A-gB & \xrightarrow{k_7} A + gB\\
B-gA & \xrightarrow{k_8} B + gA\\
S+A & \xrightarrow{k_9} S-A\\
R + B & \xrightarrow{k_{10}} R-B
\end{align*}
{\bf Design 2}\\
Here, in addition to the species in design 1, we have introduced the $P$ species, fixed to be one copy, which ensures only $A$ or $B$ can be bound at any one time
\begin{align*}
gA & \xrightarrow{k_1} gA + A \\
A  & \xrightarrow{k_2} \O{} \\
gB & \xrightarrow{k_3} gB + B \\
B & \xrightarrow{k_4} \O{} \\
A + gB + P& \xrightarrow{k_5} A-gB \\
B + gA + P& \xrightarrow{k_6} B-gA \\
A-gB & \xrightarrow{k_7} A + gB + P\\
B-gA & \xrightarrow{k_8} B + gA + P\\
S+A & \xrightarrow{k_9} S-A\\
R + B & \xrightarrow{k_{10}} R-B.
\end{align*}
{\bf Design 3}\\
Here we have the same reactions in design 1 but include two extra reactions for the decay of the bound proteins
\begin{align*}
A-gB & \xrightarrow{k_{11}} gB\\
B-gA & \xrightarrow{k_{12}} gA.
\end{align*}
{\bf Design 4}\\
Here we have the same reactions in design 1 but include two extra reactions for the binding /unbinding of the proteins $A$ and $B$
\begin{align*}
A + B & \xrightarrow{k_{11}} A-B\\
A-B & \xrightarrow{k_{12}} A + B.
\end{align*}

\subsection{Distance}
The two component distance metric was defined to be $\epsilon = \{d_1,d_2 \}$,
\begin{eqnarray}
d_1 &= & \sqrt{ \frac{ \sum_{t \in \alpha} (x_{t} - y )^2 }{ n_{\alpha} } } \nonumber \\
d_2 &= & \sqrt{ \frac{ \sum_{t \in \beta} (x_{t} - 0 )^2 }{ n_{\beta} } }, \nonumber
\end{eqnarray}
where $x_{t}$ is the number of protein $B$ at time $t$, $y$ is the target (here fixed at 40), $\alpha = \{t :  30 < t \le 60 \}$, $\beta = \{t :  0 < t \le 10 \; \text{and} \; 80 < t \le 100 \}$, $n_{\alpha} = \#\alpha$ and $n_{\beta} = \#\beta$. The final population was defined to be $\epsilon = \{7.0,0.05\}$.

\subsection{Priors}
The priors for production, binding and interaction rates were distributed as $U(0,50)$ and the priors for the degradation rates were given $U(0,5)$ distributions. 

\newpage

\begin{thebibliography}{9}
\bibitem{Martin:2003p2607}
Martin VJJ, Pitera DJ, Withers ST, Newman JD, Keasling JD (2003) Engineering a
  mevalonate pathway in escherichia coli for production of terpenoids.
  \emph{Nat Biotechnol} 21:796--802.

\bibitem{Ro:2006p2764}
Ro DK, \emph{et~al.} (2006) Production of the antimalarial drug precursor
  artemisinic acid in engineered yeast. \emph{Nature} 440:940--3.

\bibitem{You:2004p4217}
You L, Cox RS, Weiss R, Arnold FH (2004) Programmed population control by
  cell-cell communication and regulated killing. \emph{Nature} 428:868--71.

\bibitem{Kobayashi:2004p6133}
Kobayashi H, \emph{et~al.} (2004) Programmable cells: interfacing natural and
  engineered gene networks. \emph{Proc Natl Acad Sci USA} 101:8414--9.

\bibitem{Fortman:2008p2396}
Fortman JL, \emph{et~al.} (2008) Biofuel alternatives to ethanol: pumping the
  microbial well. \emph{Trends Biotechnol} 26:375--81.

\bibitem{Savage:2008p2620}
Savage DF, Way J, Silver PA (2008) Defossiling fuel: how synthetic biology can
  transform biofuel production. \emph{ACS Chem Biol} 3:13--6.

\bibitem{Cases:2005p2750}
Cases I, de~Lorenzo V (2005) Genetically modified organisms for the
  environment: stories of success and failure and what we have learned from
  them. \emph{Int Microbiol} 8:213--22.

\bibitem{Takahashi:2007p6160}
Takahashi K, \emph{et~al.} (2007) Induction of pluripotent stem cells from
  adult human fibroblasts by defined factors. \emph{Cell} 131:861--72.

\bibitem{Hanna:2010p6205}
Hanna JH, Saha K, Jaenisch R (2010) Pluripotency and cellular reprogramming:
  facts, hypotheses, unresolved issues. \emph{Cell} 143:508--25.

\bibitem{Lu:2009p6163}
Lu TK, Khalil AS, Collins JJ (2009) Next-generation synthetic gene networks.
  \emph{Nat Biotechnol} 27:1139--50.

\bibitem{Macarthur:2009p6166}
Macarthur BD, Ma'ayan A, Lemischka IR (2009) Systems biology of stem cell fate
  and cellular reprogramming. \emph{Nat Rev Mol Cell Biol} 10:672--81.

\bibitem{Anderson:2006p2634}
Anderson JC, Clarke EJ, Arkin AP, Voigt CA (2006) Environmentally controlled
  invasion of cancer cells by engineered bacteria. \emph{J Mol Biol}
  355:619--27.

\bibitem{Rajendran:2008p6134}
Rajendran M, Ellington AD (2008) Selection of fluorescent aptamer beacons that
  light up in the presence of zinc. \emph{Anal Bioanal Chem} 390:1067--75.

\bibitem{Canton:2008p2395}
Canton B, Labno A, Endy D (2008) Refinement and standardization of synthetic
  biological parts and devices. \emph{Nat Biotechnol} 26:787--93.

\bibitem{Liang:2002p6209}
Liang W, Shores MP, Bockrath M, Long JR, Park H (2002) Kondo resonance in a
  single-molecule transistor. \emph{Nature} 417:725--9.

\bibitem{Purnick:2009p2394}
Purnick PEM, Weiss R (2009) The second wave of synthetic biology: from modules
  to systems. \emph{Nat Rev Mol Cell Biol} 10:410--22.

\bibitem{Toni:2009p606}
Toni T, Welch D, Strelkowa N, Ipsen A, Stumpf MPH (2009) Approximate bayesian
  computation scheme for parameter inference and model selection in dynamical
  systems. \emph{Journal of the Royal Society Interface} 6:187--202.

\bibitem{Toni:2010p2027}
Toni T, Stumpf MPH (2010) Simulation-based model selection for dynamical
  systems in systems and population biology. \emph{Bioinformatics} 26:104--10.

\bibitem{Liepe:2010p2858}
Liepe J, \emph{et~al.} (2010) ABC-SysBio--Approximate Bayesian computation in
  Python with GPU support. \emph{Bioinformatics} 26:1797--9.

\bibitem{Bray:1994p3254}
Bray D, Lay S (1994) Computer simulated evolution of a network of
  cell-signaling molecules. \emph{Biophysical Journal} 66:972--7.

\bibitem{Francois:2004p2367}
Fran{\c c}ois P, Hakim V (2004) Design of genetic networks with specified
  functions by evolution in silico. \emph{Proc Natl Acad Sci USA} 101:580--5.

\bibitem{Battogtokh:2002p3179}
Battogtokh D, Asch DK, Case ME, Arnold J, Schuttler HB (2002) An ensemble
  method for identifying regulatory circuits with special reference to the qa
  gene cluster of neurospora crassa. \emph{Proc Natl Acad Sci USA} 99:16904--9.

\bibitem{Feng:2004p3168}
Feng XJ, \emph{et~al.} (2004) Optimizing genetic circuits by global sensitivity
  analysis. \emph{Biophysical Journal} 87:2195--202.

\bibitem{Rodrigo:2007p2375}
Rodrigo G, Carrera J, Jaramillo A (2007) Genetdes: automatic design of
  transcriptional networks. \emph{Bioinformatics} 23:1857--8.

\bibitem{Dasika:2008p2374}
Dasika MS, Maranas CD (2008) Optcircuit: an optimization based method for
  computational design of genetic circuits. \emph{BMC systems biology} 2:24.

\bibitem{Batt:2007p2392}
Batt G, Yordanov B, Weiss R, Belta C (2007) Robustness analysis and tuning of
  synthetic gene networks. \emph{Bioinformatics} 23:2415--22.

\bibitem{Ma:2009p2043}
Ma W, Trusina A, El-Samad H, Lim WA, Tang C (2009) Defining network topologies
  that can achieve biochemical adaptation. \emph{Cell} 138:760--73.

\bibitem{Elowitz:2000p2700}
Elowitz MB, Leibler S
\newblock (2000) A synthetic oscillatory network of transcriptional regulators. {\em Nature}  403:335--8.

\bibitem{Stricker:2008p1958}
Stricker J, Cookson S, Bennett MR, Mather WH, Tsimring LS, 
  Hasty J
\newblock (2008) A fast, robust and tunable synthetic gene oscillator. {\em Nature} 456:516--9.

\bibitem{Tigges:2009p2143}
Tigges M, Marquez-Lago TT, Stelling J,  Fussenegger M
\newblock (2009) A tunable synthetic mammalian oscillator. {\em Nature} 457:309--12.

\bibitem{Purcell:2010p4927}
Purcell O, Savery NJ, Grierson CS,  di~Bernardo M
\newblock (2010) A comparative analysis of synthetic genetic oscillators. {\em Journal of the Royal Society Interface} 7:1503--1524.

\bibitem{Tsai:2008p56}
Tsai TYC, Choi YS, Ma W, Pomerening JR, Tang C,  Ferrell JE
\newblock (2008) Robust, Tunable Biological Oscillations from Interlinked Positive and Negative Feedback Loops. {\em Science} 321:126--129.

\bibitem{Stock:2000p2218}
Stock AM, Robinson VL, Goudreau PN (2000) Two-component signal transduction.
  \emph{Annu Rev Biochem} 69:183--215.

\bibitem{Kim:2006p2084}
Kim JR, Cho KH (2006) The multi-step phosphorelay mechanism of unorthodox
  two-component systems in e. coli realizes ultrasensitivity to stimuli while
  maintaining robustness to noises. \emph{Comput Biol Chem} 30:438--44.

\bibitem{CsikaszNagy:2010p4748}
Csik{\'a}sz-Nagy A, Cardelli L, Soyer OS (2010) Response dynamics of
  phosphorelays suggest their potential utility in cell signalling.
  \emph{Journal of the Royal Society Interface} :Epub ahead of print.

\bibitem{Shinar:2007p2223}
Shinar G, Milo R, Mart{\'\i}nez MR, Alon U (2007) Input output robustness in
  simple bacterial signaling systems. \emph{Proc Natl Acad Sci USA}
  104:19931--5.

\bibitem{Gardner:2000p2697}
Gardner TS, Cantor CR, Collins JJ (2000) Construction of a genetic toggle
  switch in escherichia coli. \emph{Nature} 403:339--42.

\bibitem{Lipshtat:2006p5240}
Lipshtat A, Loinger A, Balaban NQ, Biham O (2006) Genetic toggle switch without
  cooperative binding. \emph{Phys Rev Lett} 96:188101.

\bibitem{Myers:2009p2142}
Myers CJ, \emph{et~al.} (2009) ibiosim: a tool for the analysis and design of
  genetic circuits. \emph{Bioinformatics} 25:2848--9.

\bibitem{Hill:2008p3637}
Hill AD, Tomshine JR, Weeding EMB, Sotiropoulos V, Kaznessis YN (2008)
  Synbioss: the synthetic biology modeling suite. \emph{Bioinformatics}
  24:2551--3.

\bibitem{Marchisio:2008p2206}
Marchisio MA, Stelling J (2008) Computational design of synthetic gene circuits1
  with composable parts. \emph{Bioinformatics} 24:1903--10.

\bibitem{baldi:1}
Baldi P, Brunak Sa (2001) \emph{{Bioinformatics: The Machine Learning Approach,
  Second Edition (Adaptive Computation and Machine Learning)}} (The MIT Press),
  2 edn.

\bibitem{Pritchard:1999p5562}
Pritchard JK, Seielstad MT, Perez-Lezaun A,  Feldman MW
\newblock (1999) Population growth of human {Y} chromosomes: a study of {Y} chromosome microsatellites. {\em Mol Biol Evol}  16:1791--8.

\bibitem{DelMoral:2006p1563}
Del Moral P, Doucet A, Jasra A (2006) Sequential Monte Carlo samplers. \emph{J. Roy. Stat. Soc. B} 68:411--436.

\bibitem{Robert:2011p7915}
Robert CP, Cornuet J-M, Marin J-M, Pillai NS (2011) Lack of confidence in ABC model choice. \emph{arXiv} 1102.4432v1.

\bibitem{Grelaud:2009p5626}
Grelaud A, Robert CP,  Marin, JM
\newblock (2009) {A}{B}{C} methods for model choice in Gibbs random fields. {\em Cr Math} 347:205--210.


\bibitem{Wilkinson:2008p3789}
Wilkinson RD (2008) Approximate Bayesian computation (ABC) gives exact results under the  assumption of model error. \emph{arXiv} 0811.3355v1.

\bibitem{Chickarmane:2005p4692}
Chickarmane V, Paladugu SR, Bergmann F,  Sauro, HM
\newblock (2005) Bifurcation discovery tool. {\em Bioinformatics} 21:3688--90.

\bibitem{Toni:2010thesis}
Toni T. (2010) Approximate Bayesian computation for parameter inference and model selection in systems biology. \emph{PhD Thesis, University of London, UK}.

\end{thebibliography}



\setcounter{figure}{0}
\renewcommand{\thefigure}{S\arabic{figure}}
\begin{figure*}[hbt]
\begin{center}
\includegraphics[width=0.95\textheight,angle=90]{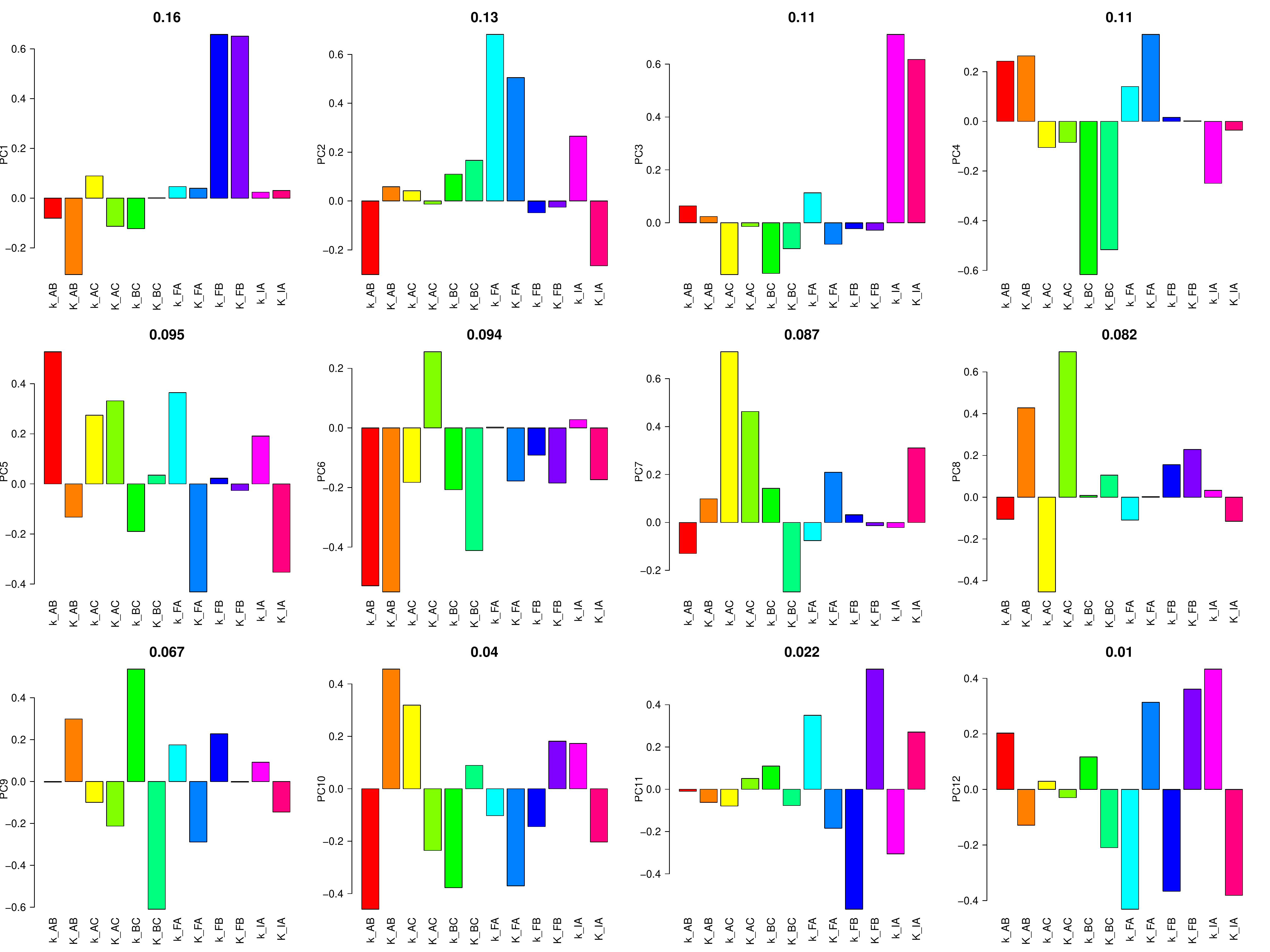}
\caption{Biochemical adaptation: principal component analysis of the posterior distribution for model 11 in the case of no cooperativity.}
\end{center}
\label{S1}
\end{figure*}

\begin{figure}[htb]
\begin{center}
\includegraphics[width=\textwidth]{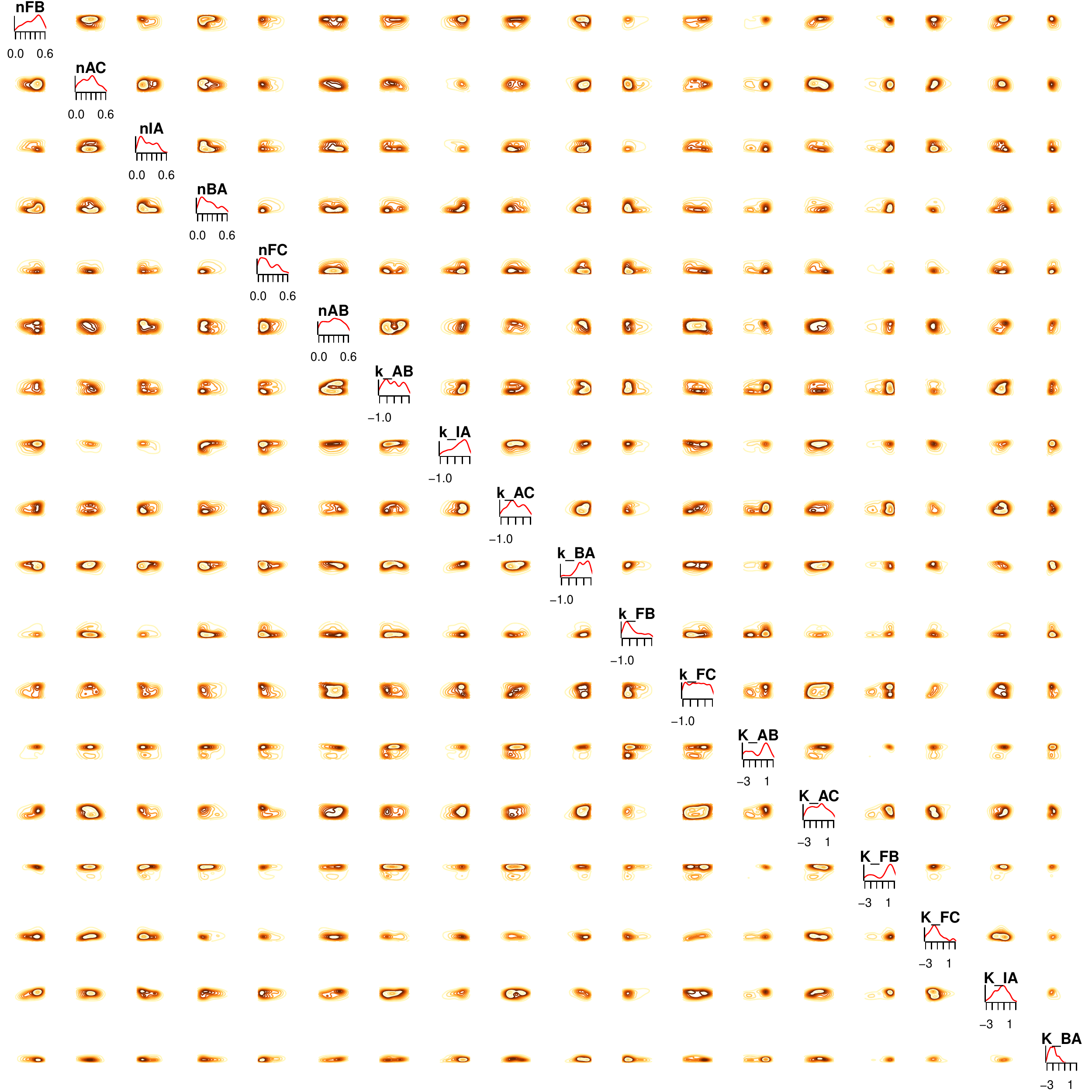}
\caption{Biochemical adaptation: posterior distribution for model 4 in the case when cooperativity is included.}
\end{center}
\label{S2}
\end{figure}

\begin{figure}[htb]
\begin{center}
\includegraphics[width=0.95\textheight,angle=90]{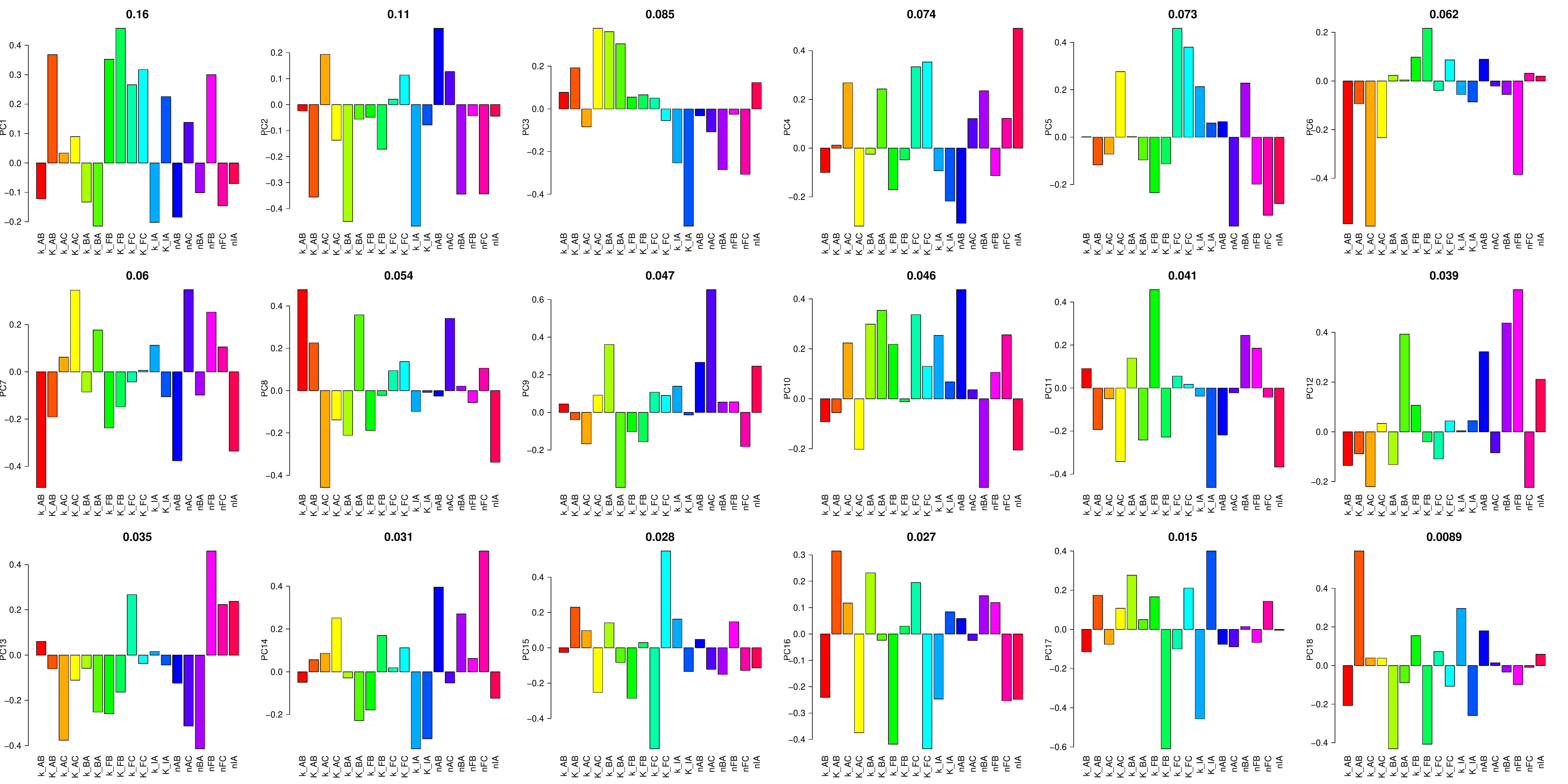}
\caption{Biochemical adaptation: principal component analysis of the posterior distribution for model 4 in the case when cooperativity is included.}
\end{center}
\label{S3}
\end{figure}


\begin{figure}[htb]
\begin{center}
\includegraphics[width=\textwidth]{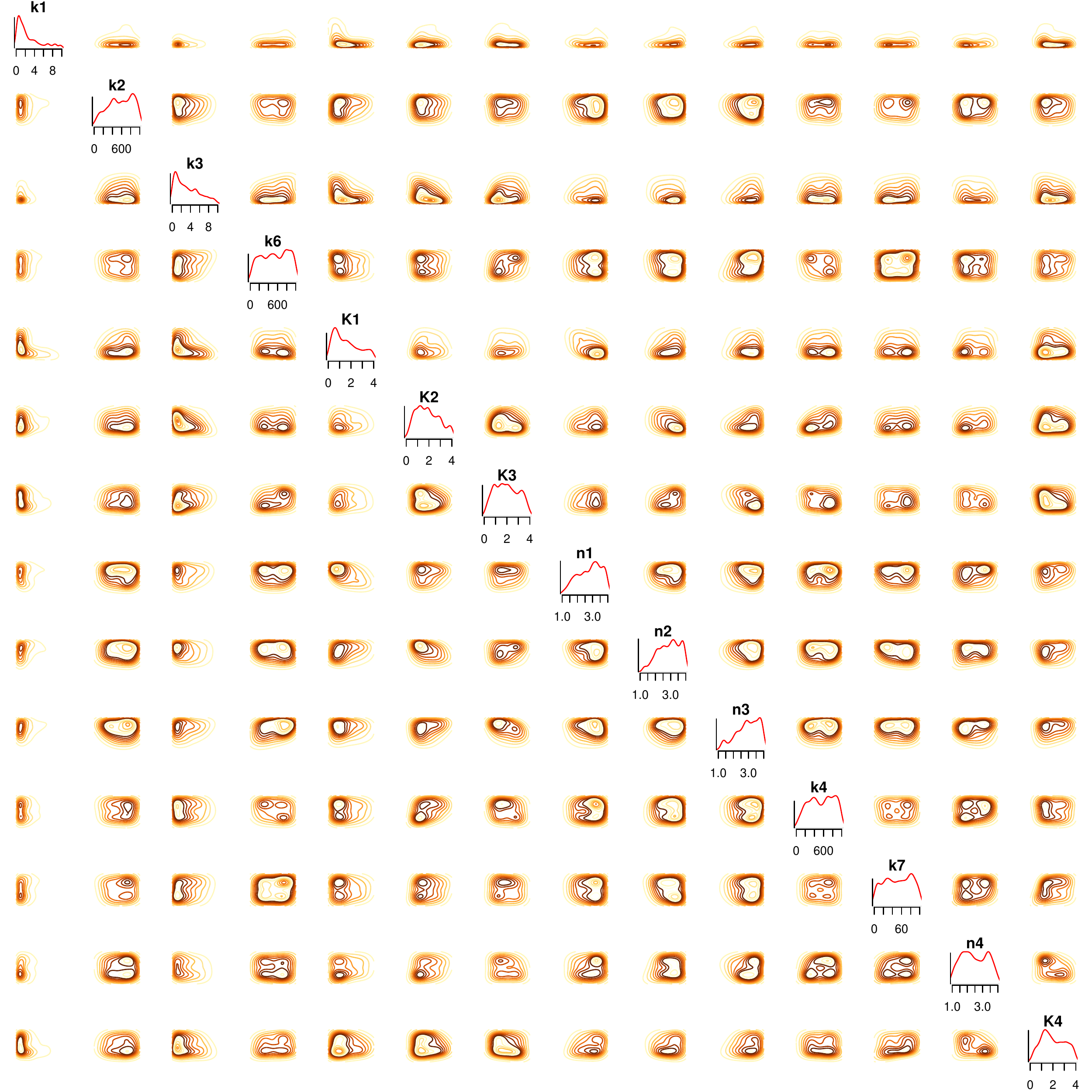}
\caption{Robust oscillator design: posterior distribution for model 2 to achieve limits cycles via a hopf bifurcation.}
\end{center}
\label{S4}
\end{figure}

\begin{figure}[htb]
\begin{center}
\includegraphics[width=0.95\textheight,angle=90]{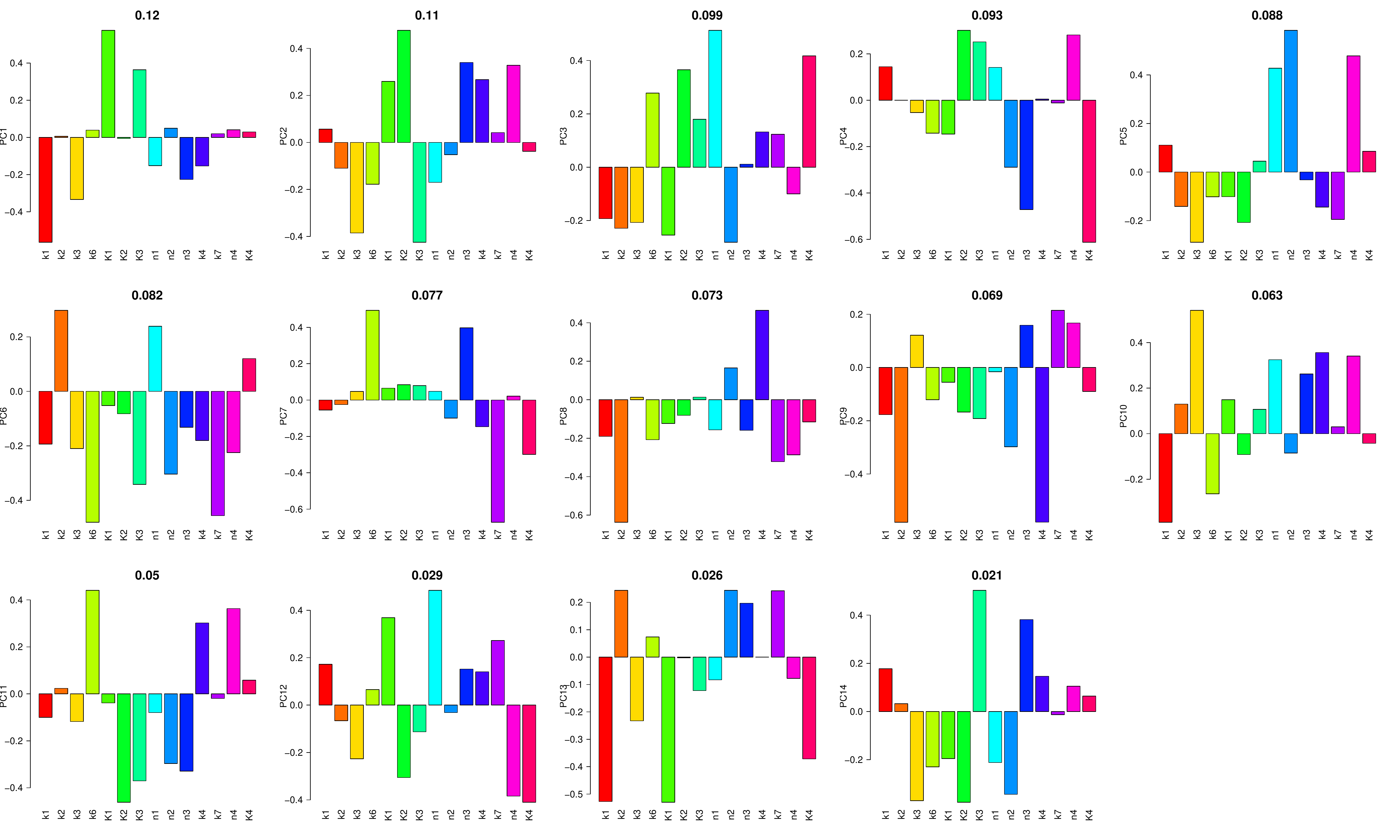}
\caption{Robust oscillator design: principal component analysis of the posterior distribution for model 2 to achieve limits cycles via a hopf bifurcation.}
\end{center}
\label{S5}
\end{figure}

\begin{figure}[htb]
\begin{center}
\includegraphics[width=\textwidth]{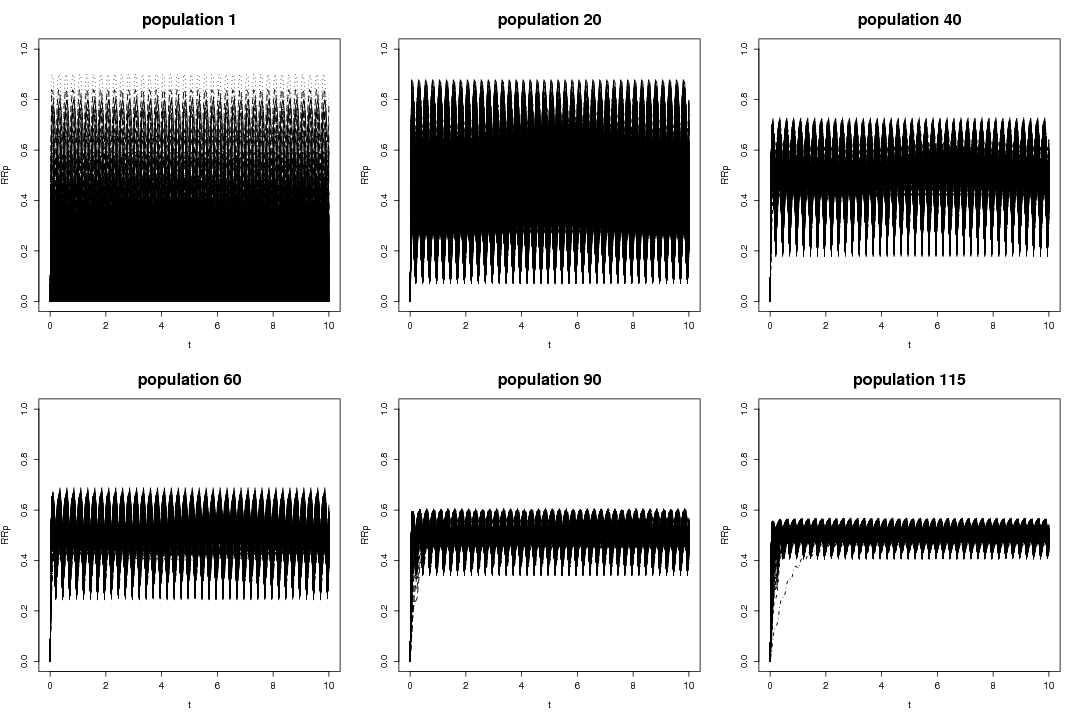}
\caption{Two component systems: evolution to the noise reduction behavior.}
\end{center}
\label{S6}
\end{figure}

\begin{figure}[htb]
\begin{center}
\includegraphics[width=\textwidth]{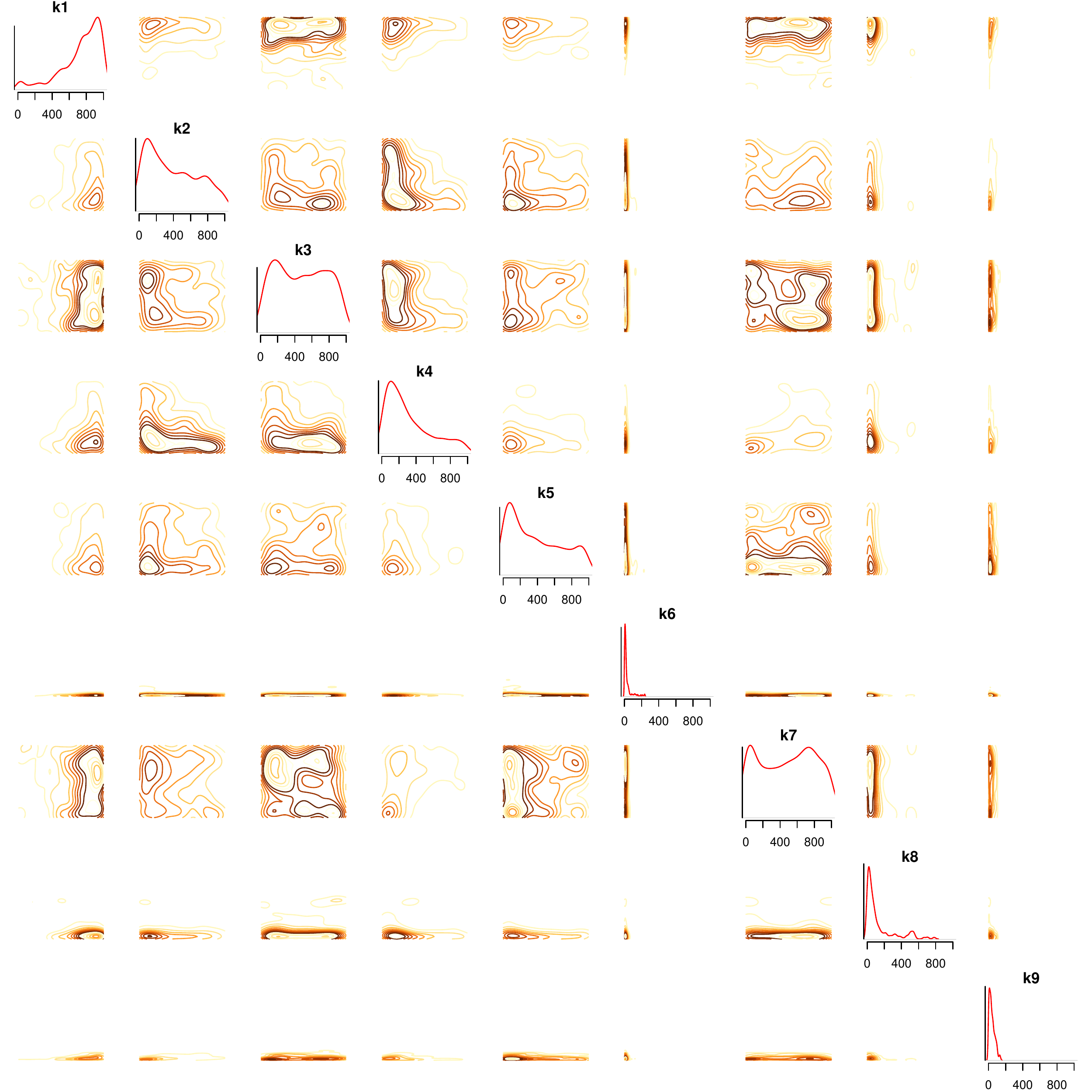}
\caption{Two component systems: posterior distribution for the unorthodox system to achieve the noise reduction behavior.}
\end{center}
\label{S7}
\end{figure}

\begin{figure}[htb]
\begin{center}
\includegraphics[width=\textwidth]{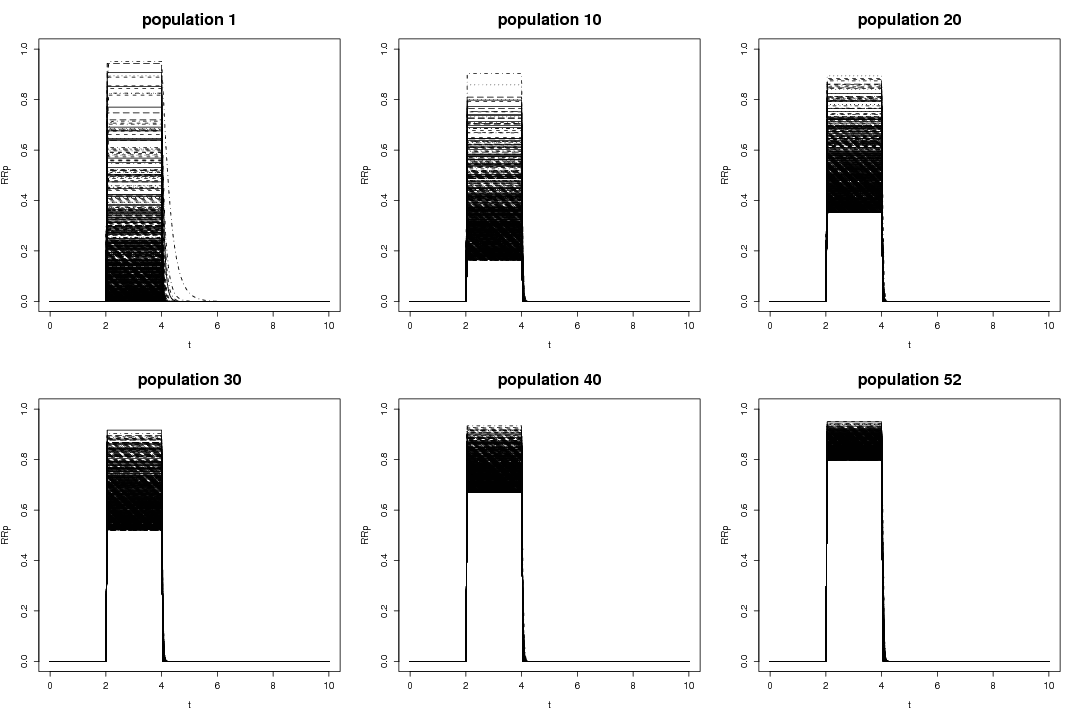}
\caption{Two component systems: evolution to the signal reproduction behavior.}
\end{center}
\label{S8}
\end{figure}

\begin{figure}[htb]
\begin{center}
\includegraphics[width=\textwidth]{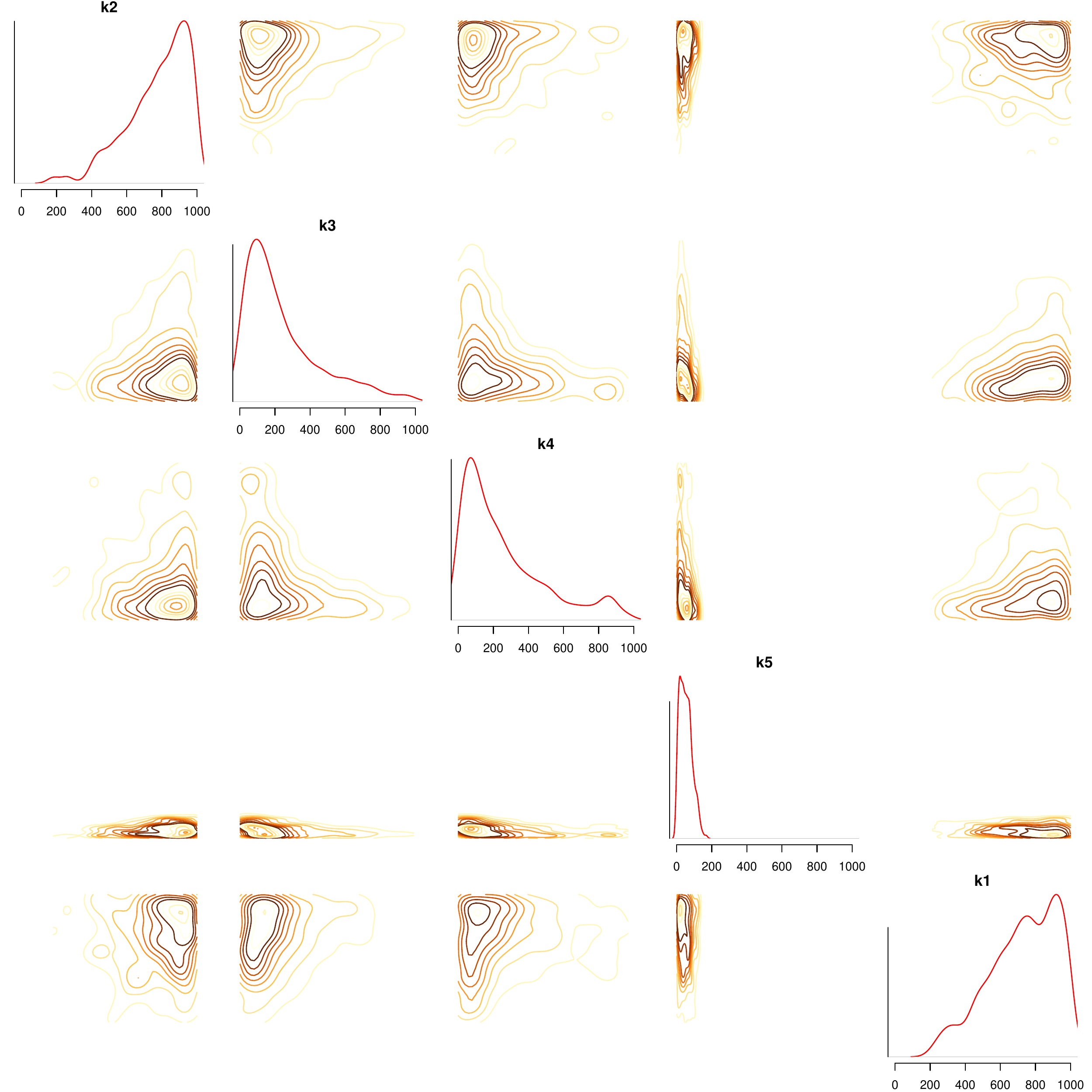}
\caption{Two component systems: posterior distribution for the orthodox system to achieve the signal reproduction behavior.}
\end{center}
\label{S9}
\end{figure}


\begin{figure}[htb]
\begin{center}
\includegraphics[width=\textwidth]{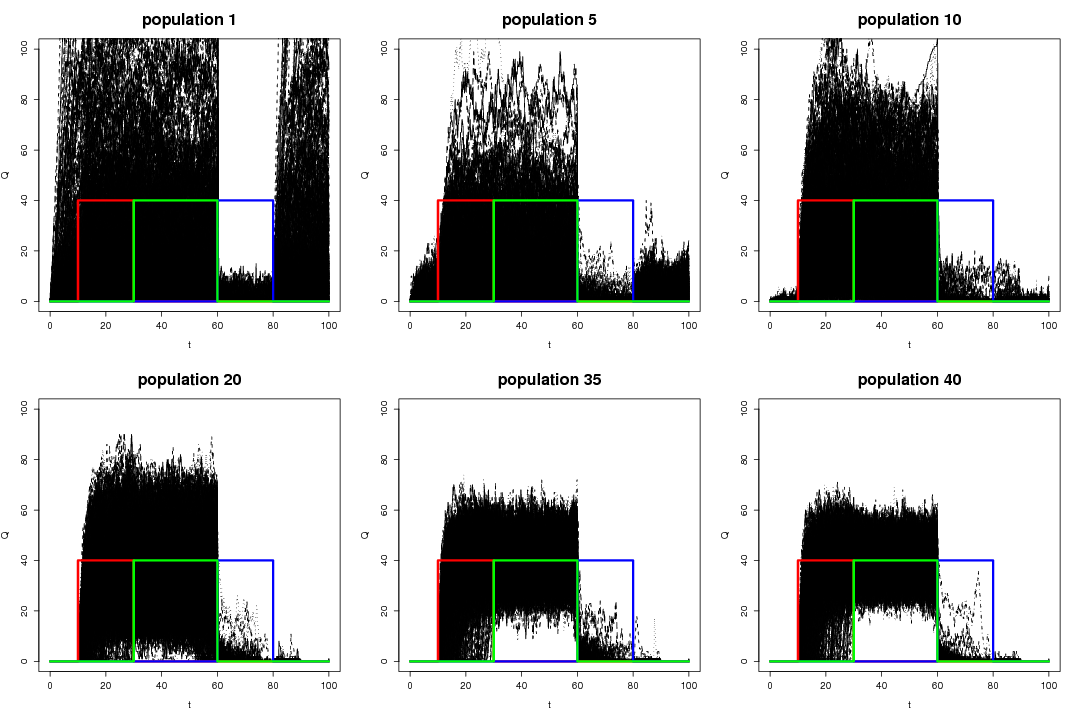}
\caption{Stochastic genetic toggle switch: evolution to the desired behavior.}
\end{center}
\label{S10}
\end{figure}


\begin{figure}[htb]
\begin{center}
\includegraphics[width=0.95\textheight,angle=90]{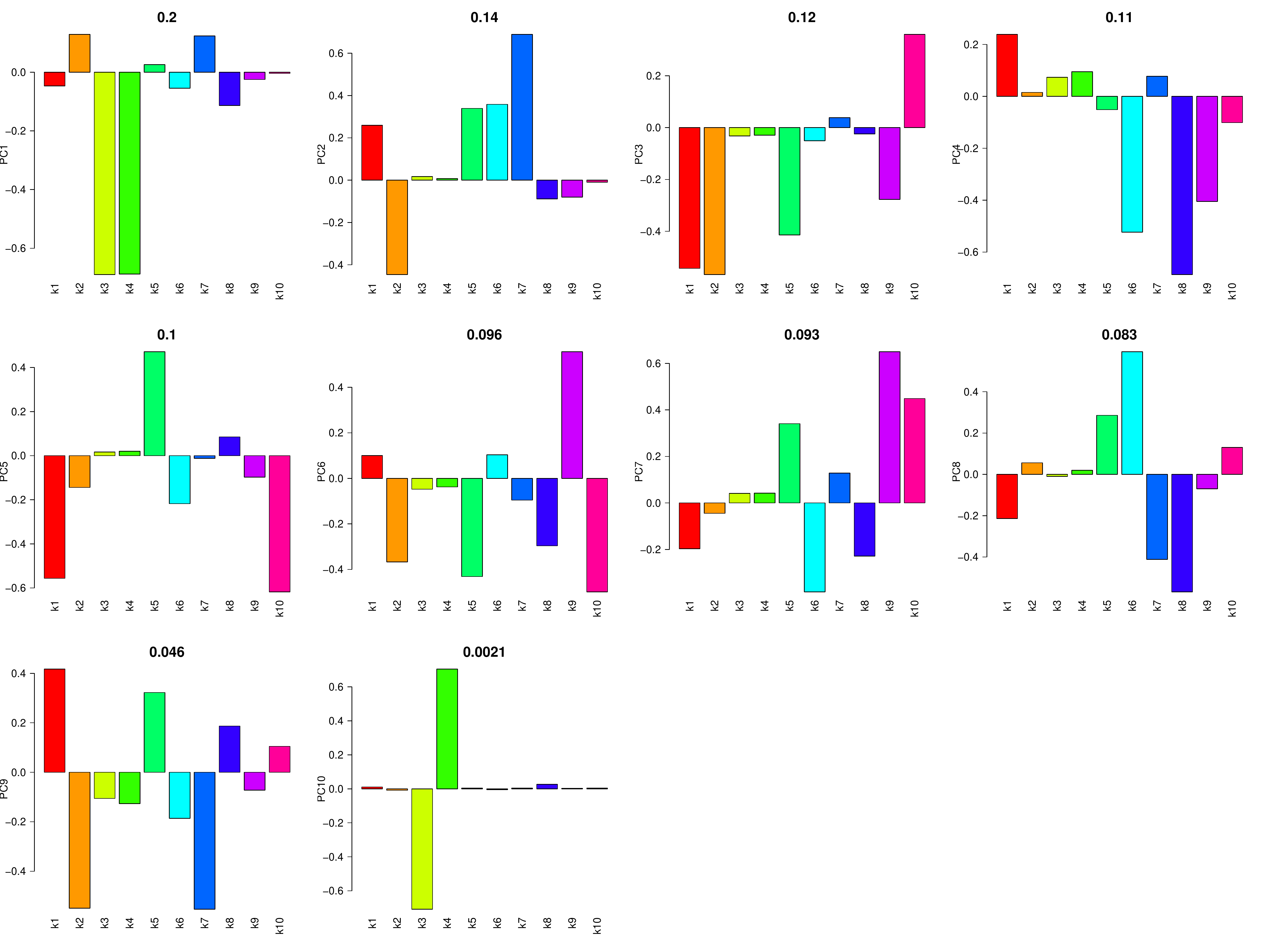}
\caption{Stochastic genetic toggle switch: principal component analysis of the posterior distribution for model 2.}
\end{center}
\label{S11}
\end{figure}

\end{document}